\documentclass[twocolumn,floatfix,groupedaddress,superscriptaddress,amsmath,a4paper,twoside,showkeys]{revtex4}
\usepackage{graphicx}
\usepackage{chemarr}
\usepackage{times,longtable}
\usepackage{hyperref,amsmath}
\usepackage{version}
\usepackage{color,pdflscape}
\usepackage{soul,amssymb}
\usepackage{longtable}
\usepackage{epsf,mathtools}
\usepackage[usenames,dvipsnames]{xcolor}

\graphicspath{{figures/},{figures/pdf/},{figures/eps/}}

\bibpunct{[}{]}{,}{n}{}{,}

\newcommand{\vcyc}{$V_{{\rm cyc}}$}

\DeclareRobustCommand{\CMRO2}{{\rm CMR}\ensuremath{_{{\rm O\ensuremath{_{{\rm 2}}}}}}}
\DeclareRobustCommand{\CMRGlc}{{\rm CMR}\ensuremath{_{{\rm Glc}}}}
\DeclareRobustCommand{\CMRGlcox}{{\rm CMR}\ensuremath{_{{\rm Glc(ox)}}}}

\providecommand{\avg}[1]{\left \langle #1 \right \rangle}

\providecommand{\addhyphen}[1]{#1.---}

\makeatletter
\renewcommand \paragraph{%
  \@startsection
    {paragraph}%
    {4}%
    {\parindent}%
    {\z@}%
    {-.1em}%
    {\normalfont\normalsize\itshape\addhyphen}%
}%
\makeatother

\begin{document}

\title{Energy metabolism and glutamate-glutamine cycle in the brain:\\A stoichiometric modeling perspective}
 

\author{Francesco A. Massucci}

\affiliation{Departament d'Enginyeria Quimica, Universitat Rovira i Virgili, 43007 Tarragona (Spain)}

\author{Mauro DiNuzzo}

\affiliation{Magnetic Resonance for Brain Investigation Lab, Enrico Fermi Center, Roma (Italy)}

\affiliation{Dipartimento di Fisica, Sapienza Universit\`a di Roma, P.le Aldo Moro 2, 00185 Roma (Italy)}

\author{Federico Giove}

\affiliation{Magnetic Resonance for Brain Investigation Lab, Enrico Fermi Center, Roma (Italy)}

\affiliation{Dipartimento di Fisica, Sapienza Universit\`a di Roma, P.le Aldo Moro 2, 00185 Roma (Italy)}

\author{Bruno Maraviglia}

\affiliation{Dipartimento di Fisica, Sapienza Universit\`a di Roma, P.le Aldo Moro 2, 00185 Roma (Italy)}

\affiliation{Fondazione Santa Lucia, Roma (Italy)}

\author{Isaac Perez Castillo}

\affiliation{Department of Mathematics, King's College London, Strand, London, WC2R 2LS (UK)}

\author{Enzo Marinari}

\altaffiliation{Authors contributed equally.}

\affiliation{Dipartimento di Fisica, Sapienza Universit\`a di Roma, P.le Aldo Moro 2, 00185 Roma (Italy)}

\affiliation{Center for Life Nano Science@Sapienza, Istituto Italiano di Tecnologia, Viale Regina Elena 291, 00161 Roma (Italy)}

\author{Andrea De Martino}

\altaffiliation{Authors contributed equally.}

\affiliation{Dipartimento di Fisica, Sapienza Universit\`a di Roma, P.le Aldo Moro 2, 00185 Roma (Italy)}

\affiliation{Center for Life Nano Science@Sapienza, Istituto Italiano di Tecnologia, Viale Regina Elena 291, 00161 Roma (Italy)}

\affiliation{IPCF-CNR, Unit\`a di Roma-Sapienza, Roma (Italy)}

\begin{abstract}
{\bf Background --} The energetics of cerebral activity critically relies on the functional and metabolic interactions between neurons and astrocytes. Important open questions include the relation between neuronal versus astrocytic energy demand, glucose uptake and intercellular lactate transfer, as well as their dependence on the level of activity.

{\bf Results --} We have developed a large-scale, constraint-based network model of the metabolic partnership between astrocytes and glutamatergic neurons that allows for a quantitative appraisal of the extent to which stoichiometry alone drives the energetics of the system. We find that the velocity of the glutamate-glutamine cycle (\vcyc) explains part of the uncoupling between glucose and oxygen utilization at increasing \vcyc\ levels. Thus, we are able to characterize different activation states in terms of the tissue oxygen-glucose index (OGI). Calculations show that glucose is taken up and metabolized according to cellular energy requirements, and that partitioning of the sugar between different cell types is not significantly affected by \vcyc. Furthermore, both the direction and magnitude of the lactate shuttle between neurons and astrocytes turn out to depend on the relative cell glucose uptake while being roughly independent of \vcyc.

{\bf Conclusions --} These findings suggest that, in absence of {\it ad hoc} activity-related constraints on neuronal and astrocytic metabolism, the glutamate-glutamine cycle does not control the relative energy demand of neurons and astrocytes, and hence their glucose uptake and lactate exchange.
\end{abstract}


\maketitle


\section{Background}

Sustained cerebral activity is crucially dependent on the functional and metabolic interplay of neurons and glial cells. Spectroscopic and imaging methods have indeed shown that the brain accommodates a wealth of cell-to-cell interactions, which ultimately have contributed to displace the decades-old notion that merely coupled whole brain activity to neuronal glucose oxidation (for a comprehensive review, see \cite{Mangia2007a}). In particular, carbohydrate metabolism is compartmentalized among neurons and astrocytes, which, together with the interstitial space, represent nearly 90\% of the tissue. Although there is evidence for the trafficking of metabolic intermediates between the two cell types, its significance and dependence on the activation state are not fully elucidated. More than 15 years ago it was hypothesized that astrocytes may support the energetics of brain function by the provision of glucose-derived lactate to neurons, in an activity-dependent manner \cite{Pellerin1994}. However, the idea of a metabolically significant astrocyte-to-neuron lactate shuttle (ANLS), as well as the activity-dependent increase in astrocytic glucose uptake, has proven to be rather difficult to confirm {\it in vivo}, while indirect and not always reproducible experimental proof was mainly obtained from experiments on cell cultures (see \cite{Dienel2006,Lak} and the excellent reviews \cite{dien,merg}).

The difficult interpretation and integration of the experimental findings produced a substantial theoretical effort aimed at characterizing intra- and inter-cellular metabolic fluxes \cite{Aubert2005,Aubert2007,Simpson2007,Mangia2009b,DiNuzzo2010c,Somersalo2012}. So far, mathematical models of transport and metabolism of glucose in neurons and astrocytes using either kinetic \cite{Aubert2005,Simpson2007,Mangia2009b,DiNuzzo2010c} or stoichiometry-based \cite{Calvetti2011,Jolivet2009,Occhipinti2009} approaches have provided conflicting results about the relevance of the cell-to-cell lactate shuttle (CCLS) (see \cite{Dienel2012a} for a recent review). This, in turn, raised some debate, especially concerning the partitioning of glucose between neurons and astrocytes and the potentially resulting intercellular lactate flow \cite{Jolivet2010}. A recent flux-balance-analysis (FBA) study indicates that the direction and magnitude of the CCLS between neurons and astrocytes depends critically on the relative uptake of blood-borne glucose \cite{casom2012}. The sharing of glucose between the two cell types is itself governed by the internal energetic demand of cells, implying that glucose partitioning alone cannot be used to draw any conclusion on the functional variations of the CCLS \cite{DiNuzzo2010c}. A critical reassessment of previous modeling results suggests that the CCLS might remain of minor significance in terms of transferred carbon equivalents \cite{Mangia2011}.

On the other hand, the known regulations of enzyme-catalyzed reactions implemented in dynamical models have so far proved insufficient to justify a fundamental energetic role for the CCLS \cite{DiNuzzo2010c}. In particular, the differences between metabolic pathways of neuronal and astrocytic networks do not imply the occurrence of lactate exchange between cells, most likely because neurons and astrocytes do possess a relatively high self-sufficiency for both glycolytic and oxidative glucose metabolism (see \cite{Dienel2012b} and references therein). This means that lactate is oxidatively metabolized in the same compartment where it is produced by glycolytic processing of glucose.

The aim of this study is to examine the activity-dependent metabolic cooperation of glutamatergic neurons and astrocytes from a network-based perspective. Specifically, two issues lie at the core of our work: (i) the correlation between partitioning of glucose and lactate shuttling; and (ii) their functional modulation across varying levels of glutamate-glutamine cycle. We have employed a constraint-based setting where an extensive and controlled sampling of the solution space is possible \cite{Martelli2009,ploscb} on a large-scale model of compartmentalized brain energy metabolism. At odds with previous studies employing constraint-based schemes to analyze the neuron-glia system in specific conditions (different from those considered here, see  \cite{Tunahan2007}), our approach does not rely on an objective function (which in our case would be hard to design) to define the relevant states. In addition, it allows to analyze in detail the feasible metabolic states for networks whose sizes are beyond those covered by other approaches like Bayesian Flux Balance Analysis (BFBA) \cite{Occhipinti2009} or Montecarlo sampling of mass-balance equations \cite{Schellenberger-RBC}. Finally, we have not made any special assumption on the regulation of biochemical pathways with respect to the activation level, nor have we imposed specific constraints on transport fluxes (except for the uptake of glucose that we use to fine-tune the oxygen-to-glucose index (OGI), see below). In short, we show that, within our stoichiometric approach: 

(a) the OGI is able to distinguish states characterized by different levels of neurotransmission, as flux configurations with larger OGIs typically carry smaller values for the velocity of the glutamate-glutamine cycle; 

(b) the partitioning of glucose between neurons and astrocytes is roughly independent of the level of activity; 

(c) the magnitude and direction of the CCLS depend strongly on glucose partitioning while being roughly independent of the level of neurotransmission. 

In other terms, within a purely stoichiometric model, the system's energetics is determined to a significant extent by the sharing of glucose. These results support the idea that neurotransmission does not impose significant constraints on glucose partitioning or CCLS. In addition, we show that (d) the overall degree of correlation among metabolic reaction fluxes between and within cells changes drastically in the presence of neurotransmission, pointing to an extended metabolic and possibly functional partnership between neurons and astrocytes.

\section{Methods}

\subsection{Network reconstruction}

The reaction network we considered (a reduced sketch of which is given in Figure~1) is composed of four main compartments: neuron (n), astrocyte (a), extracellular space (e) and blood capillary (c). 
\begin{figure*}
\begin{center}
\includegraphics[width=18cm]{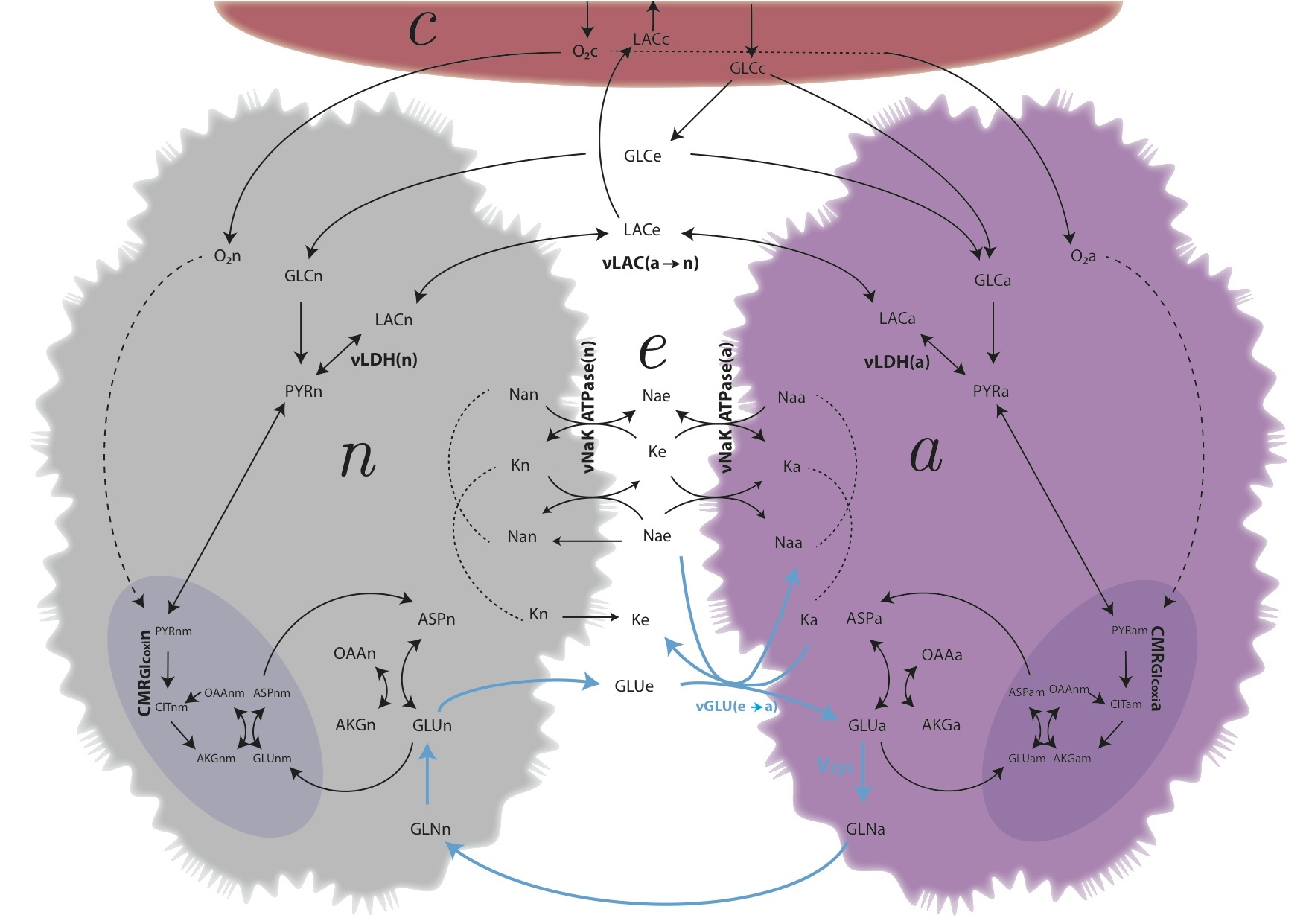}
\caption{{Schematics of the model --} The figure shows selected pathways linking the four compartments of the model (capillary, interstitium or extracellular space, neuron, astrocyte). Nutrients from the blood capillary have to traverse endothelium and basal lamina (these elements have been lumped together with the capillary) to enter the brain parenchyma. Thus, arrows connecting directly capillary and cell interior represent flows across basal lamina after endothelium. Note that this shortcut makes sense for the diffusion of oxygen to neurons and astrocytes, as well as for the transport of glucose to the astrocytic compartment only. Indeed, astrocytes but not neurons are in close apposition to cerebral blood vessels. Most of the nutrients delivery to the brain occurs through interstitial space, which is therefore the primary common element for intercellular metabolite trafficking. Once into the cells, glucose (GLC) is metabolized via glycolysis to pyruvate (PYR), which can be either reduced to lactate (LAC) or further oxidized in the cell TCA cycle requiring oxygen (O$_2$). Neuronal glutamate (GLU) is sequestered by the TCA cycle at the level of alpha-ketoglutarate (AKG) and loaded into synaptic vesicles (not shown). Neurotranmission evokes the release of vesicular glutamate into the extracellular space, from where it is taken up by astrocytes and mixed with their glutamate pool. Astrocytic glutamate can either be converted to glutamine (GLN) for export to neurons or enter the TCA cycle. The entire process consumes energy due to up-regulation of astrocytic ${\rm Na}^+/{\rm K}^+$-ATPase and glutamine synthetase ($V_{\rm cyc}$), as well as neuronal vesicle (re)filling. According to the minimal-constraints strategy employed in the present model, ionic fluxes in neurons via ligand- and voltage-gated ion channels and in astrocytes via Na$^+$/K$^+$ cotransporter follows neurotransmission passively (see text). See the Supporting Information for the full details of the network structure (139 reactions among 108 different chemical species).}
\end{center}
\end{figure*}
Within the neuronal and astrocytic elements we also distinguished the cytosol (nc and ac, respectively), mitochondria (nm and am) and synaptic vesicles (nv, only in neurons). Transport of nutrients from the blood to the brain parenchyma is provided by the capillary. We assumed that under resting conditions glucose and oxygen irreversibly enter the brain, while lactate is not significantly exchanged \cite{Hyder1998,Gjedde1997}. Specifically, glucose can be taken up directly by astrocytes via the basal lamina or can diffuse into the extracellular space \cite{Simpson2007}. The latter, in turn, is a common compartment for glucose uptake by neurons and astrocytes, as well as for lactate shuttling between the two cell types. Oxygen can freely diffuse from the capillary to cells. We lumped together the endothelium and basal lamina with the capillary compartment, which also means that we assume a negligible metabolism for endothelial cells.

The neuronal and astrocytic compartments are equipped with the enzymatic machinery to carry out the main pathways of carbohydrate metabolism (glycolysis, pentose phosphate shunt, TCA cycle, oxidative phosphorylation) \cite{Lovatt2007}. Both cell types indirectly transport reducing equivalents (i.e. NADH) from cytosol to mitochondria via malate-aspartate shuttle (MAS) (see \cite{Hertz2011}). We made the simplifying assumption that only astrocytes are capable of glutathione synthesis because neurons, unlike astrocytes, are unable to efficiently transport cystine \cite{Dringen1999} and, importantly, they cannot increase the substrate flow through glutamate-cystine ligase \cite{Gegg2003}, the rate limiting step in glutathione synthesis. Yet, the antioxidant system is equally present in neurons and astrocytes to detoxify the reactive oxygen species (ROS) produced by oxidative phosphorylation. The stoichiometry of ROS production by oxidative phosphorylation was chosen assuming that 5\% to 15\% of glucose is processed through the pentose phosphate pathway to regenerate the NADPH required for reducing oxidized glutathione \cite{Dringen1999,Gegg2003}. Anaplerosis of TCA cycle intermediates is performed by pyruvate carboxylation, which is confined to astrocytes \cite{Yu1983}, as well as by the activity of the neuronal and astrocytic malic enzyme \cite{Hassel2000}.

The functional portion of the metabolic network includes glutamatergic neurotransmission, transmitter recycling and ionic movements, that together establish the coupling between activity and metabolism through the action of the Na/K-ATPase  \cite{Ames2000}. Specifically, the glutamate stored in neuronal synaptic vesicles can be released in the extracellular space, from where it is taken up by astrocytes in co-transport with three ${\rm Na}^+$ ions and counter-transport of one ${\rm K}^+$ ion. Glutamate is amidated to glutamine by astrocytic glutamine synthetase (GS) with the concurrent hydrolysis of one molecule of ATP. Glutamine is then exported to neurons where it is eventually converted back to glutamate and loaded into synaptic vesicles again, which costs another ATP. Astrocytic uptake of glutamate and release of glutamine, together with neuronal uptake of glutamine and release of glutamate configure the so-called glutamate-glutamine cycle. In this way, the clearance of neuronally released glutamate from the extracellular space is mostly accomplished by astrocytes \cite{Danbolt2001}, although a fraction of the neurotransmitter can be taken up by neurons, especially in synapses not associated with astrocytic processes \cite{Huang2004}. At odds with previous mass-balance modeling works \cite{Occhipinti2009,Jolivet2010,casom2012}, we included the ionic currents related to membrane depolarization, albeit these were not explicitly linked to glutamate release. In particular, neurons possess ${\rm Na}^+$ and ${\rm K}^+$ channels that mimic voltage-gated ion channels and astrocytes can also take up potassium from the extracellular space with the Na-K cotransporter. Overall, the fluxes of ${\rm Na}^+$ and ${\rm K}^+$ activate  Na/K-ATPase, which consumes one ATP to transport three ${\rm Na}^+$ out of the cell and two ${\rm K}^+$ inside the cell. Importantly, not all the glutamate which is taken up by astrocytes is channeled via the glutamate-glutamine cycle. Glutamate in astrocytes can be used for energy production by entering the TCA cycle after conversion to alpha-ketoglutarate through transamination by aspartate aminotransferase (AAT) or dehydrogenation by glutamate dehydrogenase (GDH) \cite{mangiancr}. We did not include the action of other transaminase, e.g. alanine aminotransferase. This choice precludes testing the exchange of lactate and of alanine between neurons and astrocytes for maintaining ammonia homeostasis during glutamate-glutamine cycle \cite{Waagepetersen2000}. However, the role of this shuttle was experimentally found to be activity-independent in neuronal-astrocytic cultures \cite{Bak2005}. Finally, we conformed to other mass-balance modeling works \cite{Occhipinti2009,Jolivet2010} in excluding the pathways involved in the synthesis and degradation of nucleic and amino acids. This is justified by the different characteristic time-scales of processes underlying energy metabolism and gene expression, and does not rule out the possibility of any change in flux velocity brought about by e.g. protein translocation.

The network altogether consists of 139 reactions processing 108 different chemical species. The full lists of reactions and chemical species is reported in the Supporting Text.

\subsection{Flux model}

We assume that the reaction network described above operates at stationarity, i.e. that  reaction fluxes in feasible configurations are constant. More precisely, we postulate that the system is kept in a non-equilibrium steady-state (NESS) by the boundary conditions (in our case, by the fluxes of glucose and oxygen into the capillary). Although a steady-state approach for cerebral metabolism will clearly be unable to capture transient or kinetic effects, it can be justified by several considerations. In first place, a typical experiment is performed on tissue volumes containing a large number of cells, and a standard outcome will roughly represent an average over cells in the entire sample. Such averaging can be reasonably approximated with a steady--state assumption, provided the environmental conditions, including stimulation and activation, are stationary. This excludes from the analysis the time intervals associated to the transitions from one state to another (e.g. stimulation onset), which commonly last for a few tenths of a second before a steady--state is attained \cite{Mangia2007}. Related to this is the fact that in many cases the duration of a stimulus largely exceeds the equilibration time of metabolite concentrations. It has indeed been shown that sustained stimulation induces, after a short transient, a switch to different stationary states for metabolism, neuronal activity, and hemodynamic responses \cite{Bandettini1997a,Logothetis2001,Mangia2007}. Essentially, the steady-state approach allows for the study of brain metabolism on a time scale lying between the fast adaptation to the change in the activation condition and the slow adjustment of regulatory mechanisms. Finally, within this approach it is possible to treat systems much larger than those accessible to kinetic modeling (see for example \cite{Tunahan2007}), where only a few nodes of the metabolic networks are usually included.

Constraint-based models provide a standard framework for the analysis of biochemical networks in NESS. In Flux-Balance Analysis (FBA), for instance, one imposes that the vector of concentrations of intracellular metabolites $\mathbf{c}$, which in general would vary in time according to
\begin{equation} \label{fba1}
\dot{\mathbf{c}}=\mathbf{S}\boldsymbol{\nu}-\mathbf{b}~,
\end{equation}
(where $\mathbf{S}$ denotes the $M\times N$ stoichiometric matrix, $N$ is the number of reactions, $M$ that of metabolites, $\boldsymbol{\nu}$ the  vector of reaction fluxes, and $\mathbf{b}$ the vector of in- and out-takes that govern the transport of chemical species to and from the system) is constant. In turn, fluxes need to adjust to satisfy simple mass-balance conditions for the individual chemical species, amounting to the set of $M$ equations 
\begin{equation} \label{fba}
\mathbf{S}\boldsymbol{\nu}=\mathbf{b}~.
\end{equation}
Note that the elements of $\mathbf{b}$ are non-zero only for metabolites that are exchanged with the environment. For sakes of definiteness, bounds of variability for each flux $\nu_i$ ($i=1,\ldots, N$) need to be specified. Usually, such bounds  account for reversibility assignment, i.e. they are either of the form $-\infty<\nu_i<\infty$ (for reversible reactions) or of the form $0\leq \nu_i<\infty$ (for irreversible reactions), although in some cases physiological considerations may lead to consider more complicated cases, e.g. $\nu_0\leq\nu_i<\infty$ with $\nu_0>0$. (In the present study, we shall only consider bounds for the putative reversibility of reactions, as detailed in the network reconstruction reported in the Supporting Text, except for the glucose uptake flux to the capillary which is taken to be fixed. See below for details.)

The system (\ref{fba}) now defines a solution space as a polytope of dimension $N-M$ (or more precisely, $N-{\rm rank}(\mathbf{S})$; note that, typically, $N>M$). In the absence of a refinement criterion, like an {\it ad hoc} optimization prescription (see e.g. \cite{palsson0} for an excellent introduction to this modeling perspective), the set of solutions should ideally be sampled uniformly to extract both the individual solutions as well as the statistics of fluxes (averages, correlations, etc.). This is indeed the type of information we are interested in retrieving in the present case. Unluckily, exact sampling algorithms (e.g. Monte Carlo) are still inapplicable to genome-scale flux models when the dimension of the solution space exceeds a few tens because of their high computational costs \cite{recent}. (See however \cite{Braunstein:2008bf} for a promising set of alternative techniques.) In addition, the straightforward application of FBA-type of constraints in the our case is also made difficult by the fact that our reconstruction is largely incomplete. This means that the pathways we do not include may have a considerable cross-talk with the core carbon pathways on which we focus, so that the solutions of (\ref{fba}) might depend strongly on the choice of the boundary fluxes that represent the interaction between pathways included in the reconstruction and the rest of the metabolic network.

We therefore took a step back with respect to FBA and considered a broader type of conditions, inspired by Von Neumann's model of production networks \cite{vn}. In essence we simply replace (\ref{fba}) with
\begin{equation} \label{vn}
\mathbf{S}\boldsymbol{\nu}\geq \mathbf{0}
\end{equation}
for all intracellular metabolites, while keeping (\ref{fba}) for in-takes (nutrients). Clearly, the main difference with (\ref{fba}) is that (\ref{vn}) allows for flux vectors generating a net production of chemical species (corresponding to the metabolites for which the strict inequality holds in (\ref{vn})). In a nutshell, the rationale behind this is that a net production of certain compounds might be expected in cells if they need to be employed in macromolecular processes (e.g. proteinogenesis) lying out of the domain of metabolism or, more relevantly to our case, in portions of the network that are not included in the reconstruction. (In other terms, the presence of gaps and their impact on the flux organization of the core network may be smoothened out by softening the constraints.) Therefore, strictly speaking, a NESS where the concentration of certain metabolites is formally increasing in time (corresponding to the positive components of the vector $\mathbf{S}\boldsymbol{\nu}$) can be physiologically viable. Once the nutrient availability is fixed through the boundary fluxes, the cell's metabolic production and nutrient usage profiles can be determined self-consistently from the solutions of (\ref{vn}).

The solutions of (\ref{fba}) and (\ref{vn}) will obviously coincide if all inequalities in (\ref{vn}) become equalities, though in general this does not need to be the case. The main technical advantage of using (\ref{vn}) lies in the existence of an effective and scalable relaxation algorithm that allows to obtain a statistically controlled sampling of its solution space in very modest CPU times. Such a method has been defined in \cite{dmmp} and employed in e.g. \cite{Martelli2009} and \cite{ploscb} to analyze the metabolic capabilities and the energy balance of the bacterium {\it E. coli}. The statistical properties of the solution space sampling thus obtained are discussed in \cite{ploscb} and further explained in the Supporting Text. In brief, the algorithm makes use of a prior probability distribution of fluxes to initialize the flux variables and generates solutions such that  the {\it average} Euclidean distance between the solutions and the prior is minimized, the average being carried out over initial flux states sampled from the prior. A sufficiently unbiased prior (e.g. a set of uncorrelated uniform distributions, as we have employed here) then injects minimal {\it a priori} information in the solution space and therefore provides a reasonably unbiased, statistically controlled sampling of the feasible flux states of the network. In turn, such an information allows to extract physiological details of individual solutions, as well as statistical properties of the solution space (e.g. probability distributions of fluxes and correlations). This is the calculation scheme we have employed. Further details about the network (i.e. the matrix $\mathbf{S}$), the flux model and the algorithm used in the present work are found in the Supporting Text.

In the following we shall denote the flux of an intracellular reaction or of a transport process respectively by the acronym of the corresponding enzyme or the name of the transported metabolite. We shall also highlight the compartment in which the reaction occurs (so that e.g. $\nu$PDH($n$) will stand for the pyruvate dehydrogenase--catalyzed reaction taking place in the neuron) or the source/destination compartments that are involved in a transport (for instance, $\nu {\rm O_2}(c\to a)$ will denote the transport of molecular oxygen from the capillary to the astrocyte). Shorthands like ${\rm \nu O_2}(\to c)$ will instead represent the supply of metabolites (oxygen in this case) to the capillary. Unless otherwise stated, fluxes are expressed in arbitrary units and error bars correspond to one standard error.

\section{Results}

\subsection{Validation of the model: activation states}

In the network model we consider, capillaries are supplied with two compounds, namely glucose and oxygen, which can then be transferred to the other compartments. We have characterized the metabolic activity of the brain by fixing only the uptake of glucose to the capillary, ${\rm \nu Glc}(\to c)$, while leaving the oxygen influx free. In these conditions, each fixed value of ${\rm \nu Glc}(\to c)$ generates a different solution space for (\ref{vn}), where ${\rm \nu O_2}(\to c)$ fluctuates across solutions (i.e., across feasible flux configurations). This in turn yields, for each selected value of ${\rm \nu Glc}(\to c)$, a distribution of values for the oxygen-glucose index (OGI), defined as the ratio between cerebral metabolic rates of oxygen (\CMRO2) and glucose (\CMRGlc). As both nutrients do not accumulate in the tissue at steady-state, the OGI can be defined as
\begin{equation} \label{ogi}
{\rm OGI} = \frac{\CMRO2}{\CMRGlc} = \frac{{\rm \nu O_2}(\to c)}{{\rm \nu Glc}(\to c)} ~~.
\end{equation}
where nutrient influxes (both the numerator and the denominator) obey, as said above, mass-balance conditions:
\begin{gather}
{\rm \nu O_2}(\to c) = {\rm \nu O_2}(c\to a)+{\rm \nu O_2}(c\to n)\\
{\rm \nu Glc}(\to c) = {\rm \nu Glc}(c\to a)+{\rm \nu Glc}(c\to e)~~.
\end{gather}
Note that the last equation does not involve the neuronal compartment because glucose enters neurons via the extracellular space only (as said before, the endothelium and basal lamina, which mediate the transport of glucose to astrocytes and extracellular space, are included in the capillary compartment). Note also that at steady-state one has
\begin{equation}
{\rm \nu Glc}(c\to e) = {\rm \nu Glc}(e\to a)+{\rm \nu Glc}(e\to n)~~.
\end{equation}

Generically, larger OGIs imply larger fluxes through aerobic pathways, with ${\rm OGI} = 6$ as the physiologic maximum value for the steady state aerobic oxidation of glucose (corresponding to the fact that 6 oxygen molecules are required to metabolize glucose to water and carbon dioxide). Nevertheless, OGI values larger than 6 are possible in cells whenever the carbon supply for cellular respiration exceeds glucose processing through glycolysis (as happens, for instance, during lactate uptake from the bloodstream). In the flux model (\ref{vn}) it is possible to obtain OGI values slightly above 6, as a consequence of the fact that the oxygen intake to the capillary is a free variable, not bounded (within the model) by the condition ${\rm OGI}\leq 6$. From a modeling viewpoint, this may correspond to a small accumulation of intracellular oxygen due to e.g. the absence of some oxygen-consuming pathways in the network (this condition might not have a physiological counterpart). Experimental in vivo measurements show that the OGI decreases with increasing cerebral activation, from values around 5.5 (almost complete glucose oxidation) under awake resting conditions to values generally ranging from 4 to slightly above 5 during focal brain activity, depending for example on the stimulation paradigm, on the involved brain area, on the experimental procedure (reviewed in \cite{Shulman2001a,Giove2003}).  Since the contribution of individual physiologic processes under different conditions is not known, we sought to model the level of activation by using the OGI as a proxy. This assumption stems from the notion that different metabolic states can be characterized in term of their energy expenditure \cite{Shulman2011}. Furthermore, the transition to a more glycolytic than oxidative metabolism is thought to identify the transition from basal to activated conditions \cite{Raichle2002}. These arguments suggests that brain metabolism approaches full glucose oxidation as the overall signaling activity decreases.

We have hence solved (\ref{vn}) for different values of ${\rm \nu Glc}(\to c)$ recording the resulting OGI distributions. Figure~2 shows four distributions of OGI corresponding to different glucose consumption rates, starting from lower values of ${\rm \nu Glc}(\to c)$ corresponding to a larger average OGI. 
\begin{figure}
\begin{center}
\includegraphics[width=0.5\textwidth]{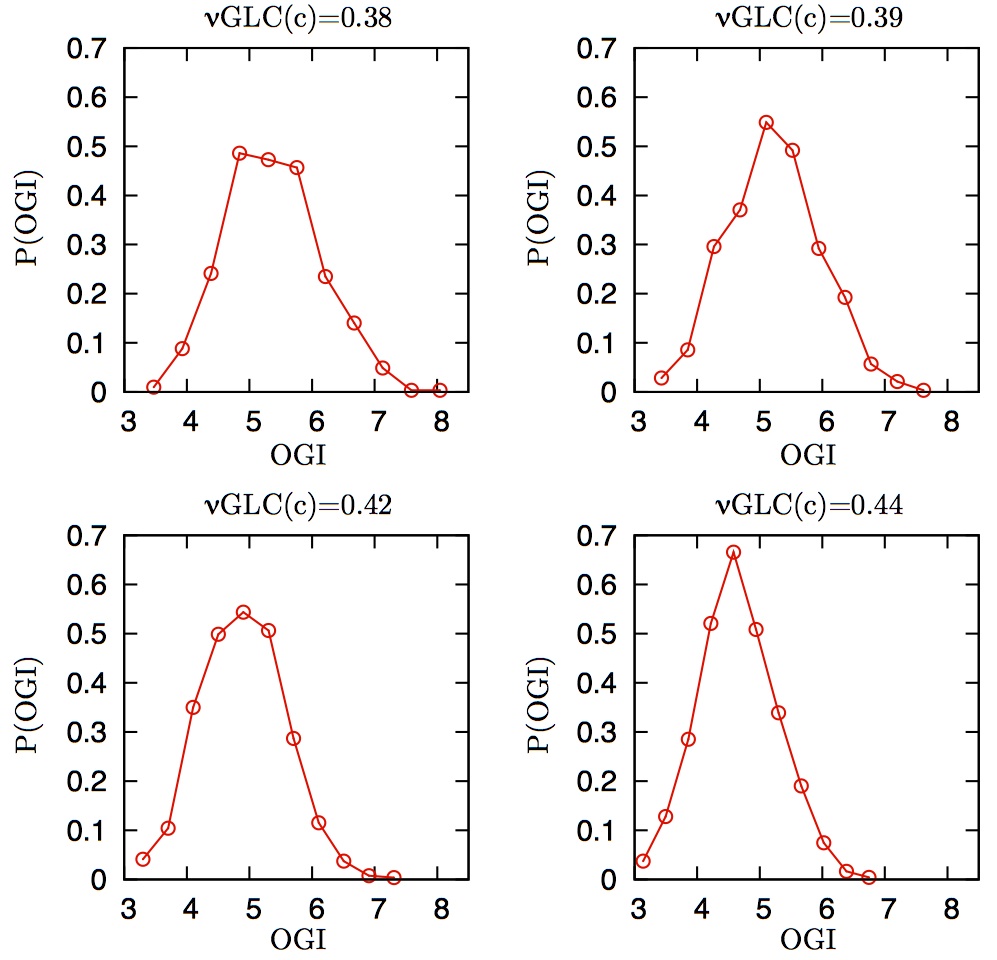}
\caption{{\bf Computed probability distributions of oxygen-to-glucose index (OGI) at different glucose uptakes --} Top left to bottom right: Uncoupling between glucose and oxygen utilization increases for increasing overall glucose uptake (i.e. enhanced glutamatergic activity). The mean OGI decreases from 5.5 to 4.5 in correspondence of an increased glucose consumption of about 15\%. This behavior is in a qualitative agreement with experimental evidence, and allows for a definition of different activation states based on the uncoupling between glucose and oxygen consumption. Each OGI distribution thus identifies a subset of solutions for subsequent flux analysis.}
\end{center}
\end{figure}
It should be noted that there is not a clear consensus about the quantitative degree of OGI decrease during activity. Indeed, while a value under activation around 5.1 is suggested by several works \cite{Shulman2001a,Giove2003}, many others point towards lower values (see e.g. \cite{Fox1986,Fox1988} and the recent \cite{Vafaee2012}). Because of this, and because the OGI distributions we found at a given ${\rm \nu Glc}(\to c)$ are rather broad, we preferred to explore a relatively broad range of OGI values.

We have further characterized the model by monitoring the degree of activation in terms of the velocity of the glutamine synthetase (GS) catalyzed reaction. In our simulated network, this univocally represents the rate of the so--called glutamate--glutamine cycle \vcyc, or
\begin{equation} \label{vcyc}
V_{\rm cyc} = {\rm \nu GS}(a)~~.
\end{equation}
It should be noted that, in vivo, there is a residual rate of Gln synthesis unrelated to neurotransmitters cycling \cite{Sibson1998}, but under physiological conditions ${\rm \nu GS}(a)\simeq V_{\rm cyc}$ \cite{Sibson1997}, and thus we conformed to our general choice of neglecting neurotransmission--unrelated amino acids synthesis also in this specific case. The assumption that the residual rate of Gln synthesis is independent of glutamatergic neurotransmission is made also in the original experimental work that reported $V_{\rm cyc}$ \cite{Sibson1998}, and the relevant implications are discussed therein.

As happens for the OGI, each choice of ${\rm \nu Glc}(\to c)$ leads to a distribution of values of $V_{\rm cyc}$. Generically, solutions with $V_{\rm cyc}=0$ (i.e. no neurotransmission \footnote{Because of finite numerical precision, for our purposes a flux is null when it is smaller than $10^{-6}$.}) will coexist with solutions carrying a non-zero level of activation for any choice of ${\rm \nu Glc}(\to c)$. In order to highlight the quantitative changes induced by activation, we performed a correlation analysis between the fluctuations of flow rates of reaction or transport pathways, first in all the sampled solutions (i.e. with $V_{\rm cyc}\geq 0$) and then in solutions with $V_{\rm cyc}>0$. It turns out (see Figure 3) that the transition from uncostrained neurotransmission to presence of neurotransmission has profound consequences on the flux correlations within and between cells.
\begin{figure*}
\begin{center}
\includegraphics[width=18cm]{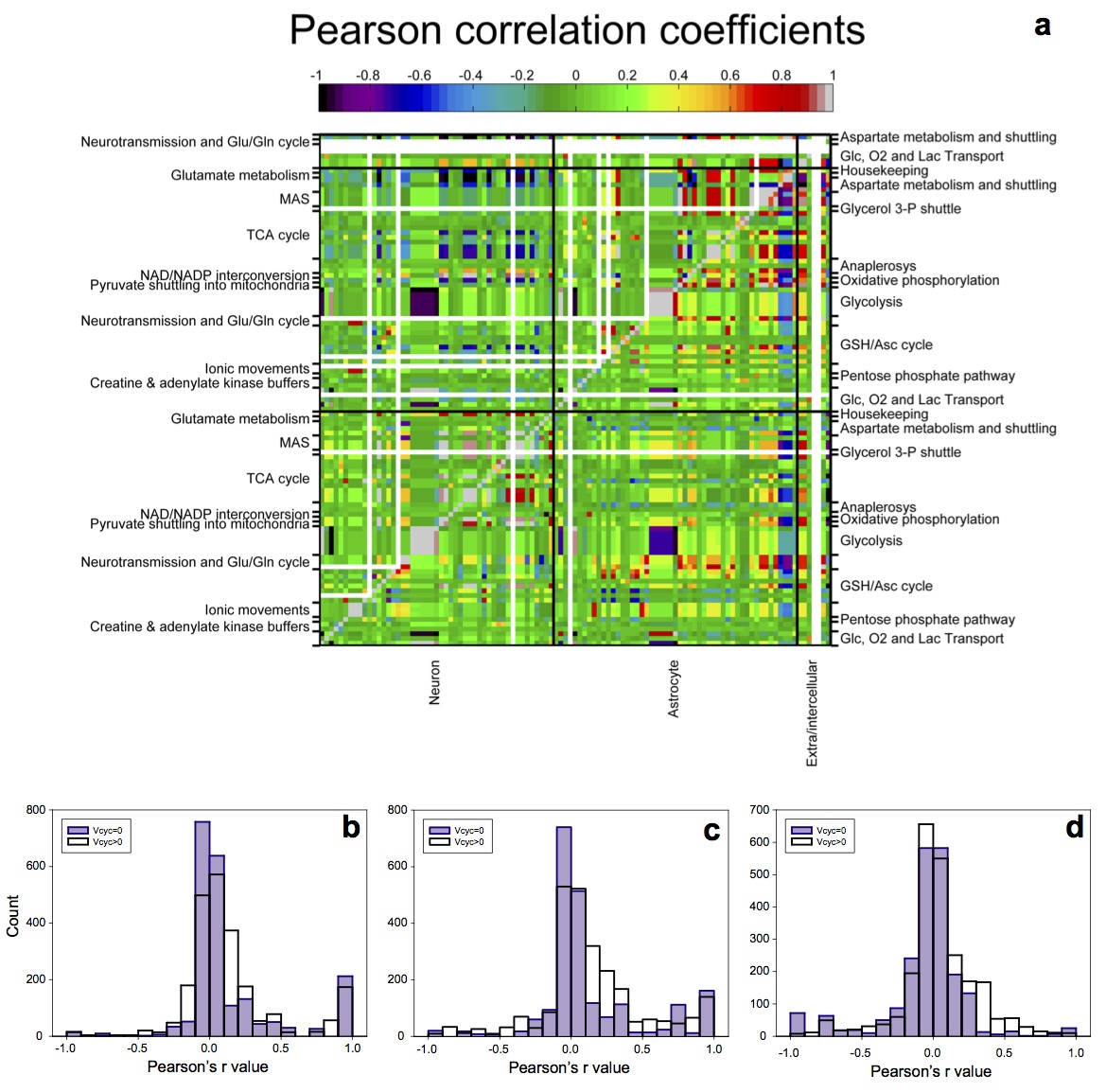}
\caption{{\bf Pearson correlation coefficients for each pair of reactions -- }{\bf (a)} Cycles, pathways and homogeneous reaction groups are reported sideways, the compartment is reported at the bottom. Neuron and astrocyte sector also include transports from/to the cell, while the extra/intercellular class groups reactions that either directly connect neuron to astrocyte, or involve only extracellular compartments. Null fluxes are represented in white. Above the diagonal we report the Pearson's coefficients obtained by imposing $V_{\rm cyc}=0$, while below the diagonal $V_{\rm cyc}$ is allowed to assume positive values. {\bf (b,c,d)} Histograms of the Pearson correlation coefficients for each pair of reactions within neurons (b), astrocytes (c) or between cells (d). For each plot, the distributions are reported for $V_{\rm cyc}=0$ (in pale blue) and $V_{\rm cyc}>0$ (transparent). Null fluxes (i.e. those fluxes that are plotted in white in panel (a)) are excluded from the histograms. Apart of the apparent larger number of null fluxes with $V_{\rm cyc}=0$, the histograms show that the bins at higher correlation tend to be more populated at $V_{\rm cyc}> 0$ than at $V_{\rm cyc}=0$, while the opposite holds for central bins (i.e. those bins with very low direct or inverse correlation).}
\end{center}
\end{figure*}

The solutions obtained for $V_{\rm cyc}>0$ underlie a substantial increase of both the global and the regional correlation level of the network. This is especially true for the correlation between neuronal and astrocytic metabolism, confirming that the condition $V_{\rm cyc}>0$ identifies the functional and metabolic partnership between the two cell types. This finding shows that intercellular signaling has a major role for the overall metabolic regulation at tissue level, constraining the catabolism of each cell in a concerted range (see Discussion).

The above results suggest that, in order to capture the quantitative changes that occur in solutions at higher levels of activation (recall that our model does not, per se, constrain the magnitude of neurotransmission), it is useful to analyze the behavior of the conditional average $V_{\rm cyc}$ versus the conditional average OGI.
In Figure~4 we display the results obtained by retaining solutions with $V_{\rm cyc}\geq 0$ (i.e. all solutions, returning the standard average) and $V_{\rm cyc}>0$, respectively. 
\begin{figure}
\begin{center}
\includegraphics[width=0.5\textwidth]{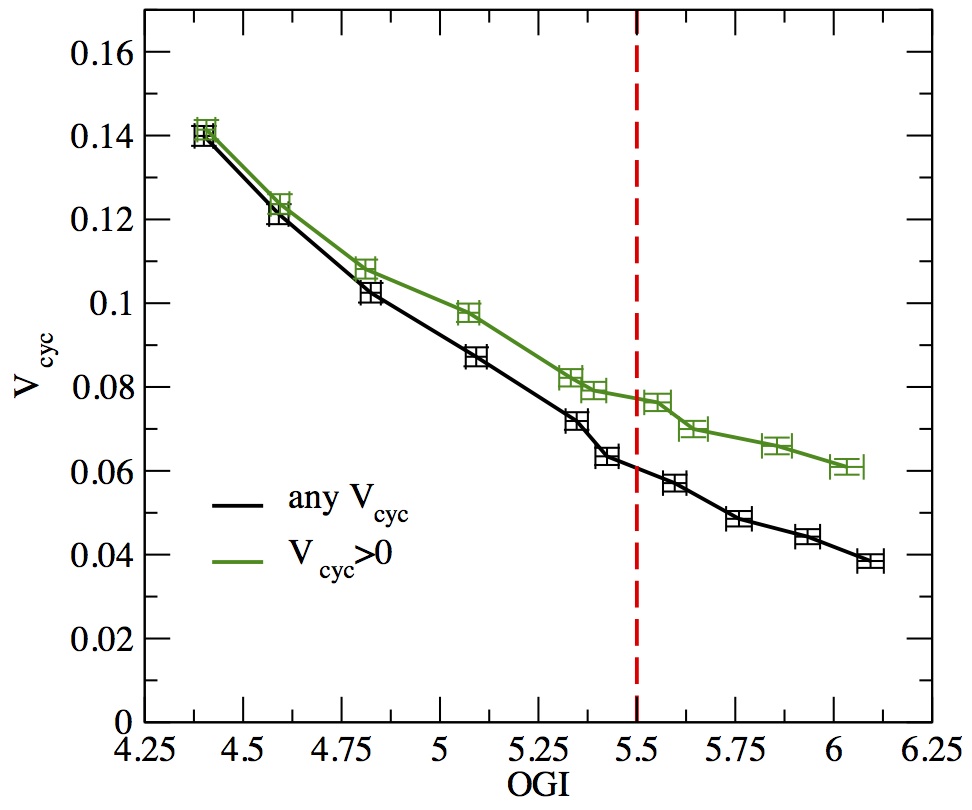}
\caption{{\bf Computed average rate of the glutamate-glutamine cycle versus average OGI -- }Simulations show that the average $V_{\rm cyc}$ increases as the average OGI decreases, consistently with the underlying relationship between OGI and glucose uptake. The curves represent conditional averages, computed over solutions characterized respectively by $V_{\rm cyc}\geq 0$, and $V_{\rm cyc}>0$. The red dashed line identifies the awake resting state (OGI $=5.5$). The crossing points between the line at OGI $= 5.5$ and each of the two curves identify the values of $V_{\rm cyc}$ relevant for the basal conditions. Specifically, $V_{\rm cyc}$ is roughly 0.06 or 0.08 for the groups $V_{\rm cyc}\geq 0$ and $V_{\rm cyc}>0$, respectively. The two curves are significantly different for OGI $\gtrsim 4.75$. This is consistent with the fact that the contribution of solutions with $V_{\rm cyc}=0$ becomes negligible at high activation levels.}
\end{center}
\end{figure}

One sees that as the average OGI decreases, the average $V_{\rm cyc}$ increases to an enhanced activation level. The slope of the curve is less negative for the condition $V_{\rm cyc}>0$, i.e. when the sample is restricted by filtering--out the states with suppressed neurotransmission. This changes the basal state at ${\rm OGI} = 5.5$ to a rate of glutamate-glutamine cycle larger by about 30\%, from 0.06 to 0.08 (roughly).

In summary, in agreement with the literature, we find that the rise in anaerobic glucose consumption during activation turns out to be out of proportion to oxygen utilization, as evidenced by the decrease in the (average) OGI for increasing values of \vcyc\ and \CMRGlc\ . This indicates that the glutamate-glutamine cycle by itself suffices to explain part of the uncoupling between glucose and oxygen utilization. It should be emphasized that the reported OGI reduction, although potentially significant if caused by a specific subset of glycolytically-served energetic demands, does not change the overall strategy of brain energy metabolism, which remains largely aerobic because of the higher ATP yield of respiration \cite{Mangia2007a}.

\subsection{Neuronal oxidative glucose metabolism versus glutamatergic activity}

The relationship between glutamate-glutamine cycle and neuronal glucose oxidation was experimentally reported to be close to a 1:1 relation \cite{Hyder2006}. In particular, the rate of neuronal oxidative metabolism of glucose (i.e. the level of activity of pyruvate dehydrogenase, PDH) increases linearly with the velocity of transmitter cycling, with a slope close to one. Our framework allows to address the dependence of glucose oxidation in neurons on the rate of glutamate-glutamine cycle for various states of activation. Considering that PDH is the primary entry point of glucose-derived pyruvate into the TCA cycle, we define
\begin{equation} \label{CmrOxi}
{\CMRGlcox}(n) =\frac{1}{2} {\rm \nu PDH}(n)  \, ,
\end{equation}
where the factor $1/2$ is required as glycolysis produces two molecules of pyruvate for each glucose molecule. Plotting the average $\CMRGlcox(n)$ against the average $V_{\rm cyc}$ we find approximately two different regimes around the physiologic range corresponding to OGI $\simeq 5.5$ (see Figure 5). 
\begin{figure}
\begin{center}
\includegraphics[width=0.5\textwidth]{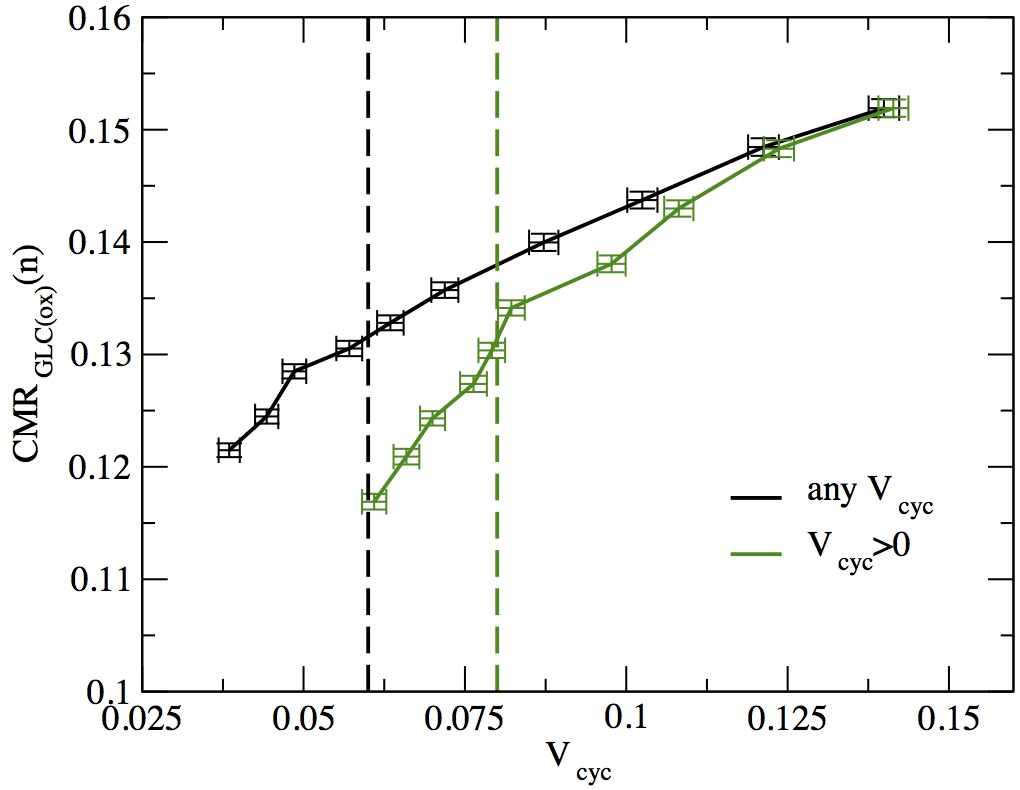}
\caption{{\bf Rate of glucose oxidation in neurons as a function of the Glu/Gln cycle --} The curves represent conditional averages, computed over solutions characterized respectively by $V_{\rm cyc}\geq 0$ and $V_{\rm cyc}>0$.  The  dashed lines identify the awake resting state (OGI $=5.5$), corresponding to the group of solutions ($V_{\rm cyc}\geq 0$ and $V_{\rm cyc}>0$) plotted with the same color. For the condition $V_{\rm cyc}>0$, and possibly also for the condition $V_{\rm cyc}\geq 0$, the awake rest roughly corresponds to a change of the line slope. Thus, the transition from low to high activity is accompanied by a decreased energy consumption relative to what would be extrapolated at low activity levels. In particular, the slopes of the two curves at low $V_{\rm cyc}$ values are roughly 0.52 and 0.73 (for the groups $V_{\rm cyc}\geq 0$ and $V_{\rm cyc}>0$, respectively) and decrease to roughly 0.28 and 0.33 at high $V_{\rm cyc}$ values, suggesting a lack of energy demand at high activity (see text).}
\end{center}
\end{figure}
These regimes are characterized by almost linear relations, in agreement with the experimentally reported constant stoichiometry between aerobic Glc oxidation in neurons and glutamate cycling \cite{Sibson1998,Hyder2006}.  However, the coupling pattern  changes as one explores states with values of $V_{\rm cyc}$ departing from the basal level to higher activity. The slope of the curve clearly increases if only strictly positive $V_{\rm cyc}$ are considered, although it stays below one. Interestingly, in the high $V_{\rm cyc}$ regime (upstream the awake value, for which no experimental data exist) the plot features a slight attenuation of the curve. The fact that the slope is lower than the linear extrapolation at high activity indicates that some source of energy consumption adds to the glutamate--glutamine cycle during the transition from normal to high neurotransmission levels. In this region, the discrepancy between simulated and experimental data is explained by the relatively low increase in ionic fluxes through neuronal voltage-gated $\text{Na}^+$ and $\text{K}^+$ channels obtained in our simulations (data not shown), suggesting that the glutamate--glutamine cycle alone is insufficient to account for the rise in brain glucose utilization. To this end, it is mandatory for stoichiometric models to incorporate energy use by action and synaptic potentials in addition to glutamatergic neurotransmission.

\subsection{Modulation of glucose uptake and cell-to-cell lactate shuttling by glutamate-glutamine cycle}

The fate of carbons undergoing oxidative phosphorylation in neurons and astrocytes provides a quantitative hint of the relative amount of energy produced by aerobic pathways in the two cell types. We found that anaerobic and aerobic metabolism is similarly increased in both cell types at increasing activity (not shown). In particular, the fraction of cerebral oxidative metabolism in astrocytes is about 35\% of the total, which is consistent with a substantial astrocytic contribution to functional brain energy metabolism (reviewed in \cite{Hertz2007}).

Analysis of glucose fluxes showed large fluctuations, so that both the neuron and the astrocyte may be the primary sites of glucose consumption at fixed values of Glu/Gln cycling. We indeed observed that both cell compartments can absorb from 10\% to 90\% of the total glucose uptake respectively. Therefore, at this level of detail, the network does not place significant constraints on the cellular utilization of glucose. The failure of up-regulation of ionic fluxes that we report here might play a role if their contribution is substantially different for neurons and astrocytes, which unfortunately has not been yet experimentally determined.

The direction of the shuttle of lactate depends in a robust way on the sharing of glucose between neurons and astrocytes, resulting in ANLS when the relative astrocytic glucose uptake becomes larger than about 65\% (implying that states supporting ANLS can be obtained by an {\it ad hoc} adjustment of Glc partitioning) and NALS otherwise (Figure 6). 
\begin{figure}
\begin{center}
\includegraphics[width=0.5\textwidth]{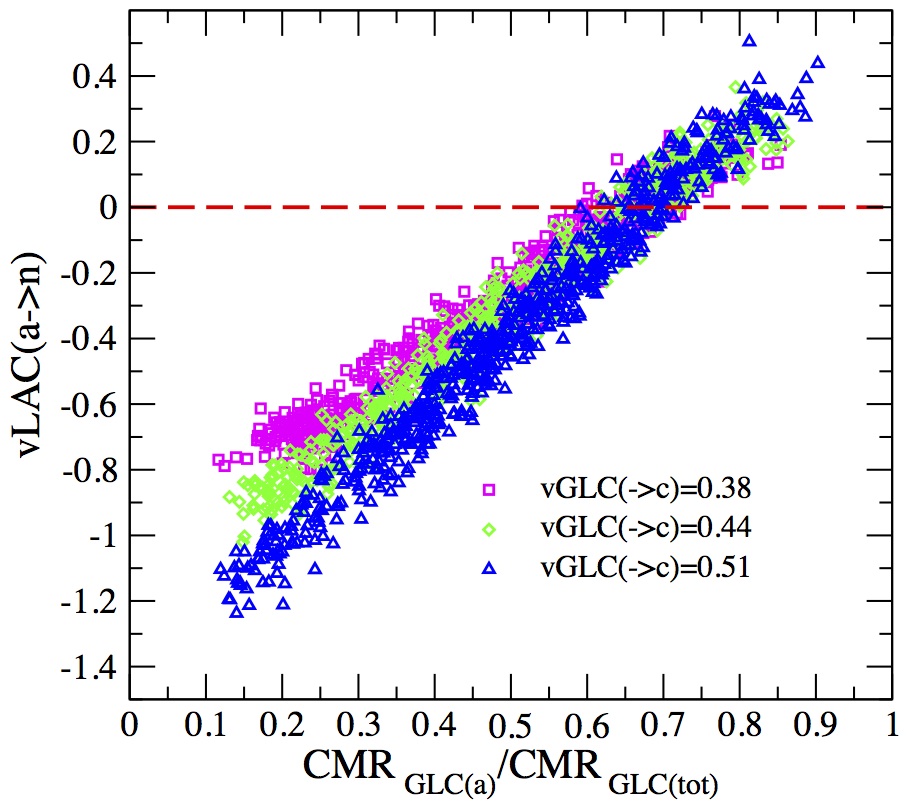}
\caption{{\bf Intercellular lactate flow versus glucose partitioning between neuron and astrocyte -- }There is a clear dependence of the CCLS on the cell glucose uptake. However, when partitioning of glucose between neurons and astrocytes is around 65\% (note that ${\rm CMR}_{{\rm Glc(tot)}}={\rm CMR}_{{\rm Glc(a)}}+{\rm CMR}_{{\rm Glc(n)}}$), which also identifies the concomitant fraction of oxygen utilization, the contribution of transferred lactate is minimal. This means that he pyruvate derived from CCLS is thus always much less than the pyruvate generated by the concomitant uptake of glucose. Notably, if glucose is taken up equally by the neuronal and astrocytic compartments, the direction of lactate flow is preferentially from neurons to astrocytes ($\nu {\rm LAC} (a\to n)\simeq -0.3$), contributing on average about 40\% of the total carbons metabolized by these cells (note that carbons from lactate are obtained by considering the halved value of the flux).}
\end{center}
\end{figure}
As the latter is also the mean value for the fraction of neuronal versus astrocytic oxygen utilization, it turns out that, on average, the contribution of CCLS to cell metabolism is very low compared to the lactate generated intracellularly by glucose. The strict dependence of CCLS on cellular glucose uptake supports previous results obtained through various modeling approaches \cite{DiNuzzo2010c,Mangia2011,casom2012}.

A closer look at the individual solutions reveals the absence of a significant correlation between the glutamate--glutamine cycle and both the uptake of glucose and the shuttle of lactate (Figure 7). 
\begin{figure}
\begin{center}
\includegraphics[width=0.5\textwidth]{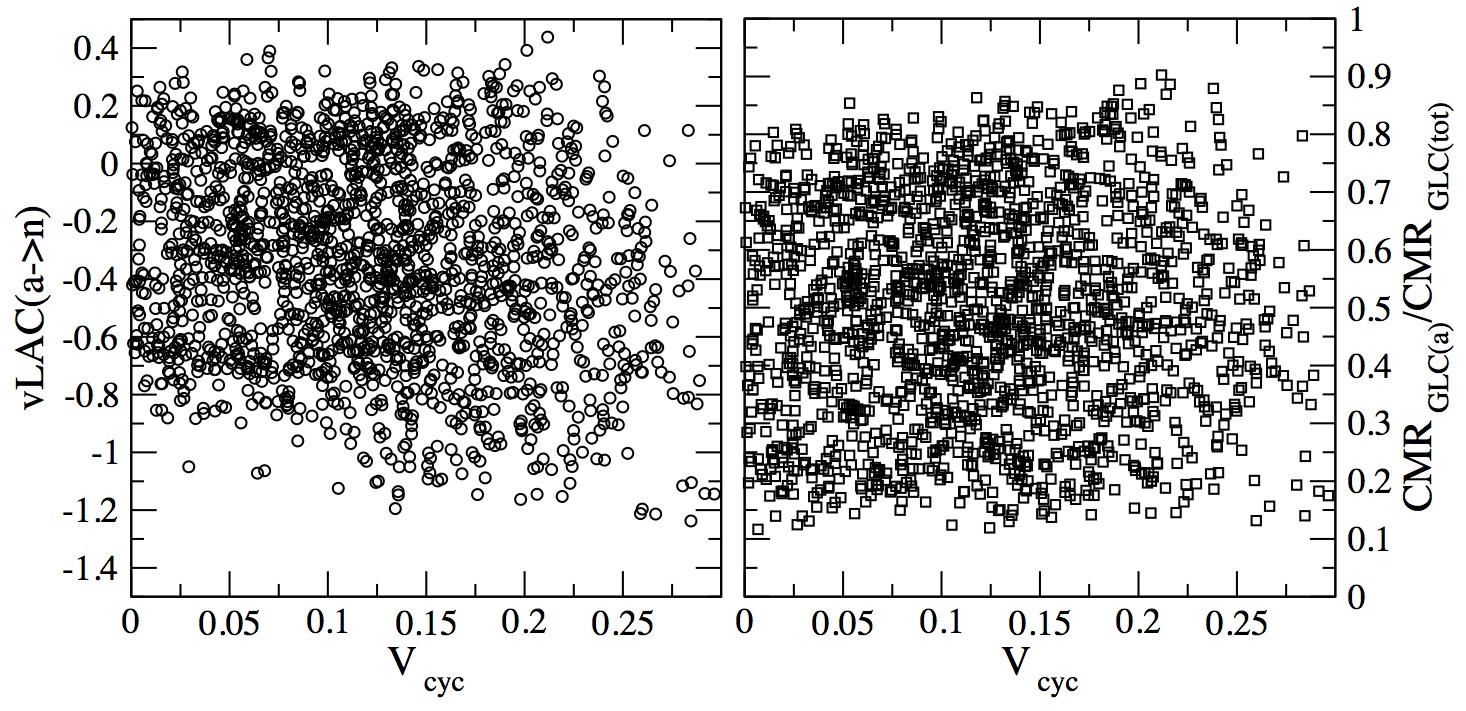}
\caption{{\bf Net ANLS flux (left) and relative cell glucose uptake (right) versus the velocity of the Glu/Gln cycle -- }Independently of the correlation between $\nu {\rm LAC} (a\to n)$ and glucose partitioning observed in Fig.~5, neither variable shows significant correlation with the activation level. This results from the absence of constraints imposed by stoichiometry on the rate of neurotransmission (hence on the glutamate-glutamine cycle), the latter being compatible with a large set of solutions relative to glucose partitioning and lactate shuttling.}
\end{center}
\end{figure}
It should however be stressed that, as can be seen in Figure 7, the emerging scenario presents large fluctuations, in the sense that, even within the physiological range for the OGI, solutions with ANLS and NALS coexist. Together, these simulation outcomes indicate that no preferential route is undertaken by lactate at any given rate of glutamate-glutamine cycle. Thus, the determinants for lactate accumulation and shuttling, if any, must reside elsewhere, for example in the balance between spiking and synaptic activity \cite{yetanother}.

\subsection{Conserved moieties, transcellular aspartate shuttling and glutathione-ascorbate cycle}

To conclude, we discuss the flux organization of several pathways that were included in the metabolic network reconstruction, but whose involvement is not strictly related to lactate trafficking. The ATP buffering systems of creatine and adenylate kinase in neurons and astrocytes, as well as astrocytic glycogen metabolism were found to have negligible net flux, consistently with the role of adenylates, creatine and glycogen as conserved moieties in the steady-state. We are unable to test hypotheses about the role of brain glycogen \cite{Shulman2001,DiNuzzo2010d,DiNuzzo2011}, because our approach does not allow to describe the consumption of previously stored metabolites, such as glycogen in this case. We found that the contribution to NADH shuttling by the malate-aspartate shuttle (MAS) dominates over the glycerol-3-phosphate pathway (relative flux is a few percent). We could not support the recent hypothesis of a significant steady state neuron-to-astrocyte transport of aspartate \cite{Pardo2011}. However, this finding was expected due to the presence of mitochondrial aspartate-glutamate carrier in astrocytes, which allows for a shunting pathway, alternative to neuron-to-astrocyte aspartate transport that is sufficient to sustain NADH shuttling from cytosol to mitochondria in astrocytes, as previously suggested \cite{Hertz2011}. In order to examine a possible role of (dehydro)ascorbate transfer between neurons and astrocytes \cite{Castro2009}, we included the detoxification of reactive oxygen species (ROS) via glutathione/ascorbate cycle. Unluckily, we were unable to find statistically significant exchange fluxes of the two forms of ascorbic acids between cells. This is possibly correlated with the fact that, as said above, we did not include the entire network section associated with glutathione synthesis. Theoretical analysis of transcellular ascorbate cycling will thus require further refinements of the network reconstruction.

\section{Discussion}

This work is concerned with a model of brain energy metabolism consisting of four compartments, representing neuron, astrocyte, extracellular space and the capillary. We used a steady state approach based on Von Neumann's theoretical framework for the analysis of production networks \cite{vn}, successfully applied to cellular metabolism in previous studies \cite{Martelli2009}. The steady-state assumption for the metabolic coupling between oxygen and glucose consumption underlying different cortical states is consistent with (i) the fact that experimental parameters have been measured under stationary conditions during suppressed brain activity \cite{Sibson1998}, and (ii) the establishment of a new metabolic steady-state during enhanced brain activity \cite{Mangia2007}. At odds with standard schemes based on flux-balance analysis though, the frame employed here doesn't constrain the net metabolite production to zero, but, rather, aims at recovering the steady state self-consistently from minimal stability requirements. The `soft' type of constraints thus arising makes it possible to sample the solution space corresponding to our large-scale network model in a statistically controlled manner, and returns a full range of feasible values for each reaction flux in the network, as well as detailed information on correlations. Based on this, one can elucidate the extent to which stoichiometry alone constrains the operation of brain metabolism, since the emerging picture is obtained without imposing specific functional constraints.

First, we found that the rate of glutamate-glutamine cycle distinguishes different activation states according to the fraction of glucose that is processed via glycolysis versus respiration. In particular, decreasing OGI values predict increasing velocity of the cycle, and viceversa. 
Notably, ATP-consuming Na$^+$ and K$^+$ fluxes across voltage- and ligand-gated ion channels are not directly dependent on glutamate-glutamine cycle in our theoretical account. Pathways analysis showed that their reaction rates do not ``automatically'' up-regulate along with transmitter cycling. This indicates that the absence of causal changes (i.e. constraints) in these different aspects of neuronal signaling strongly underestimates the glucose utilization at high activity levels. Unfortunately, it is presently unknown how to model the exact cause-and-effect relationship between glutamate-glutamine cycle and the ionic currents that generated neurotransmission on one hand, and those that are generated by neurotransmission on the other. The latter will constitute a primary target of future studies.

Second, the results of the present model support previous kinetic analyses indicating that lactate derived from astrocytes may not provide an important source of carbon compounds for neuronal energy metabolism in an activation-dependent manner (Figure 6) \cite{DiNuzzo2010c}. The model also supports the conclusion that the direction and magnitude of CCLS are secondary to glucose partitioning between cell types \cite{DiNuzzo2010c,Mangia2011}. We found that glucose uptake, glycolysis and respiration change proportionally in a wide range of glutamate recycling rates. This implies that the glucose taken up by each cell is completely metabolized and not significantly converted to lactate. Thus, lactate does not accumulate in a specific cell type, i.e. the lactate concentration gradient, and hence lactate shuttle, remains small.

Third, simulations showed that glutamate-glutamine cycle is correlated with overall tissue glucose utilization and lactate production, but not with specific patterns of cellular glucose uptake and lactate shuttle (Figure 7). 
These results agree well with the experimental knowledge \cite{Bak2009, Sibson1998} and recover features of different (and sometimes conflicting) numerical studies performed previously \cite{Jolivet2010,Mangia2011}. Most importantly, they add further arguments to the idea that the trafficking of molecules between neurons and astrocytes underlies a broad functional, rather than strictly energetic partnership (see, e.g. \cite{Bergersen2012}). This is evident in our modeling perspective, as possibly energy-related changes in lactate fluxes are independent of concomitant function-related variations in glutamate-glutamine cycle (see Figure 7, 
that essentially shows a lack of correlation between lactate shuttling and transmitter cycling). Alternative functions for lactate include the discrimination between dilation and constriction of cerebral arterioles \cite{Gordon2008}, or the modulation of GABAergic inhibitory activity of specific neuronal populations \cite{Shimizu2007}. These functional roles of lactate might be still secondary to its accumulation in the tissue. Our simulations suggest that this accumulation likely results from up-regulation of non-oxidative metabolism in both neurons and astrocytes, as previously suggested \cite{DiNuzzo2010c,DiNuzzo2010d}.

We found that intercellular shuttling of lactate increases only when glucose partitioning between neurons and astrocytes is significantly uneven (Figure 6). 
Therefore, lactate transfer can be interpreted as a local biochemical shunt that allows for the optimal use of carbon supply in correspondence of variable environmental challenges. 
Interestingly, if the functional partnership between neurons and astrocytes is suppressed via the zeroing of neurotransmission, the anticorrelation between glycolysis in the two cellular compartments further increases (i.e. correlation becomes more negative), a feature shared with many other fluxes related to energetics, including TCA cycle and oxidative phosphorylation. So, even when there is no functional relationship between neurons and astrocytes, their energy (primarily glycolytic) metabolism is anticorrelated. The reason for this behavior is that the two cell types share the same environment, thereby the same glucose availability. Glucose availability thus represents the primary drive on glucose partitioning. If the two cells are functionally independent, oxidative metabolism follows glycolysis, as also evidenced by the negative correlation between neuronal and astrocytic oxidative metabolism.  As soon as neurons and astrocytes become coupled by neurotransmission, the network shifts to correlated patterns of activity, both within the same compartment and between different compartments. Accordingly, glycolysis in one cell type becomes slightly positively correlated with oxidative metabolism in the other. However, glycolysis in either neurons or astrocytes, which essentially reflect partitioning of tissue glucose, remains poorly correlated with the glutamate-glutamine cycle. Overall, glutamate-glutamine cycle positively correlates with cell TCA cycle and oxidative phosphorylation much more than it does with glycolysis, which is especially significant for the astrocytic compartment.

In conclusion, we developed a large-scale model  for compartmentalized brain energy metabolism including the core carbon pathways of neurons and astrocytes, as well as further compartments (capillary, extracellular space). A constraint-based scheme was then employed in order to define feasible configurations of reaction fluxes. Our numerical analysis was based on a relaxation algorithm which allows to obtain a statistically controlled sampling of the solution space without any prior assumption on the behaviour of energy producing/consuming pathways. Results have shown that only a large imbalance of cell glucose uptake can explain the occurrence of a significant lactate shuttle between neurons and astrocytes, the latter being roughly independent of the rates of transmitter cycling. Our results therefore do not support a link between glutamate-glutamine cycle and CCLS as a mechanism for activation-dependent transfer of energy compounds within the brain. On the other hand, CCLS can be found by assuming that neurons have limited access to glucose and/or by bounding their glucose uptake flux. The lack of correlation we observe stems from the fact that the distribution of activity-related energy stress between neurons and astrocytes cannot be estimated by the stoichiometry of the metabolic network and should be a primary target of current experimental research. Future developments will focus on introducing minimal {\it ad hoc} constraints on neurotransmitter cycling and ionic fluxes in the hope to capture the ``non-stoichiometric'' side of the energetics of brain activity.

\section*{Acknowledgments}

This work is supported by the DREAM Seed Project of the Italian Institute of Technology (IIT) and by the joint IIT/Sapienza Lab ``Nanomedicine''. The IIT Platform ``Computation'' is gratefully acknowledged. FAM acknowledges financial support from European Union Grants PIRG-GA-2010-277166 and PIRG-GA-2010-268342.

\appendix

\vspace{9cm}

\centerline{{\bf SUPPORTING TEXT}}

\section{Constraint-based flux model}

The standard approach to modeling biochemical reaction networks is based on enzyme kinetics and is normally formulated through systems of differential equations for the time evolution of the intracellular concentrations of metabolites, like
\begin{equation}\label{dynn}
\dot{\mathbf{c}}=\mathbf{S}\boldsymbol{\nu}-\mathbf{u}
\end{equation}
where $\mathbf{c}$ denotes the concentration vector, $\mathbf{S}$ the matrix of stoichiometric indices, $\mathbf{v}$ the flux vector (in turn, a function of concentrations, kinetic constants $\mathbf{k}$, etc.: $\boldsymbol{\nu}\equiv\boldsymbol{\nu}(\mathbf{c},\mathbf{k},\ldots)$), and $\mathbf{u}$ the vector of uptakes describing the exchange of chemical species with the environment. Limited knowledge of enzyme mechanisms and kinetic constants as well as the fast increase in the number of parameters, however, effectively circumscribes this framework's applicability to small systems (up to a few tens of reactions as in the case of human erythrocites \cite{Jamshidi:2001kl}). Large- (possibly genome-) scale reaction networks require, to date, a different, simplified type of analysis.

Constraint based models are usually defined by the requirement that in non equilibrium steady states (NESS) the net rate of change of the level of metabolites is zero (i.e. by homeostasis). As a consequence, (\ref{dynn}) reduces to
\begin{equation}\label{mb}
\mathbf{S}\boldsymbol{\nu}=\mathbf{u}~~~,
\end{equation}
and one is interested in retrieving the flux patterns $\mathbf{v}$ that are consistent with a NESS induced by the (given) vector $\mathbf{u}$ of boundary fluxes \cite{Varma:1994kc,kau,Lee:2006qo, palsson,Beard:2008fk,Oberhardt:2009gb,Orth:2010if}. In essence, constraint-based models retain the information encoded in $\mathbf{S}$ as the key input and use it to define, through (\ref{mb}), a polytope of dimension $D=N-\text{rank}(\mathbf{S})$ ($N$ being the number of reactions in the network) that, if complemented with physiological bounds on the flux variables and with detailed prescriptions for in- and out-takes, can provide a sensible representation of the metabolic capabilities of the cell in a given medium. In addition, they allow for the straightforward integration of biochemical, empirical or thermodynamic data (when available) into the problem, either as specific bounds on fluxes (e.g. for reaction directionality) or in the form of extra constraints (e.g. conservation laws). In the present study, each flux $\nu_i$ ($i=1,\ldots,N$) is assumed to be either irreversible, in which case its bound of variability is simply $0\leq\nu_i<\infty$, or reversible, in which case $-\infty<\nu_i<\infty$. In the latter case, however, we introduce two irreversible fluxes to describe the forward and reverse processes respectively, so that, once every reversible reaction is split in two, all bounds we consider are of the form $0\leq\nu_i<\infty$ (except for the glucose uptake flux to the capillary which, as explained in the main text, is assumed to be fixed.)

For many microbial metabolic networks $\mathbf{S}$ is known with gene-level accuracy. In such cases, a characterization of the cell's metabolism can be obtained by sampling (ideally with uniform probability) the space of feasible network configurations defined by (\ref{mb}). Unluckily, the task of generating solutions uniformly out of the polytope is computationally unaffordable when $D$ becomes larger than a few tens (for typical genome-scale networks, $D$ can be as large as several hundreds) \cite{Price:2004lp,Braunstein:2008bf}. For systems like bacteria, however, it is possible to reduce the complexity of the solution space by coupling (\ref{mb}) with the optimization of a (usually linear) score function representing the biological functionality of the organism in the selected extracellular conditions (e.g. biomass flux maximization under optimal growth conditions for E. coli)  \cite{Varma:1994gd,Edwards:2002bd,Feist:2008bs,Lewis:2010fu}. This type of approach provides a further unquestionable advantage in terms of computational tractability and genome-scale models of metabolism have been developed along these lines, for several single-cell organisms.

For cells carrying no clear objective function the latter approach is much harder to justify and sampling solutions appears as the most logical step to take. A viable alternative to (\ref{mb}) in such cases consists in relaxing the mass-balance constraint to allow for a net production of chemical species, while leaving the network functionally unconstrained so that the metabolite production profile can be determined self-consistently. This scenario is described by the system of inequalities
\begin{equation}\label{vn2}
\mathbf{S}\boldsymbol{\nu}\geq\mathbf{0}~~~,
\end{equation}
which are easily obtained from (\ref{mb}) if one includes in- and out-takes in $\mathbf{S}$ (and the corresponding fluxes in $\mathbf{v}$). Steady-state conditions like (\ref{vn2}) were originally introduced by Von Neumann in the analysis of input-output networks \cite{gale} and simply state that, in non equilibrium steady states, flux configurations in which the network produces a metabolite in excess of consumption are allowed. In particular, solving (\ref{vn2}) for $\boldsymbol{\nu}$ after fixing a configuration of in-takes (out-takes being in this case an outcome) allows to retrieve an $M$-dimensional vector $\mathbf{y}=\mathbf{S}\boldsymbol{\nu}$ whose entries encode information on whether metabolite $j$ is being produced ($y_j>0$) or not ($y_j=0$) in that particular solution \cite{Imielinski:2005dz,ib00}.

The physiological rationale to employ such constraints is two-fold. On one hand, a net production of certain chemical species (e.g. amino acids) must be expected to take place if macromolecular processes outside metabolism strictly defined (e.g. proteinogenesis) are to occur. In absence of an objective function that accounts for such processes, (\ref{vn2}) appear as a reasonable minimal constraints for the metabolic capabilities of a cell, in the sense that they don't even impose which chemical species are to be globally produced. On the other hand, when the network reconstruction is incomplete one should complement (\ref{mb}) with additional constraints that account for the flow of chemical species to network modules not included in the model. Such out-takes may be hard to implement in absence of detailed genomic data. Applying (\ref{vn2}) to partial network reconstructions allows to deal with this issue rather naturally, by exploring all possible metabolic exchanges compatible with the given stoichiometry. Moreover, its dual problem has recently been given a thermodynamic interpretation in the context of cell metabolism \cite{Warren:2007fk,ploscb}.

\section{Numerical analysis: relaxation method}

The flux problem (\ref{vn2}) has been studied from a purely theoretical perspective in \cite{Martino:2005mi,Martino:2007zt,Martino:2009lh,Martino:2010tw}. The technical advantage that accompanies (\ref{vn2}) lies in the existence of a computationally efficient,  statistically controlled method to generate solutions. The procedure consists in essence of a Relaxation algorithm \cite{linopt} based on the analogy of (\ref{vn2}) with perceptron learning \cite{minover0} and is described in detail in \cite{Martino:2007zt}. In brief, denoting by $\mathbf{A}$ and $\mathbf{B}$ the matrices of input and output stoichiometric indices respectively (so that $\mathbf{S}=\mathbf{B}-\mathbf{A}$), let $\rho>0$ be a real parameter, let $\mathbf{S}_\rho=\mathbf{B}-\rho\mathbf{A}$, and consider the system
\begin{equation}\label{vn1}
\mathbf{S}_\rho\boldsymbol{\nu}\geq\mathbf{0}~~.
\end{equation}
Given a (generic) flux vector $\boldsymbol{\nu}$ (e.g. randomly generated from a prescribed probability distribution), a solution of (\ref{vn1}) can be found for any fixed $\rho<1$ by the following algorithm:
\begin{itemize}
\item compute $\mathbf{y}=\mathbf{S}_\rho\boldsymbol{\nu}$ and $j_0=\text{arg} \min_j y_j$  (i.e., $j_0$ is the index of the least satisfied constraint);
\item if $y_{j_0}\geq 0$ then $\boldsymbol{\nu}$ is a solution of (\ref{vn1}); exit.
\item if $y_{j_0}< 0$ then update $\boldsymbol{\nu}$ component-wise as
\begin{equation}\label{up}
\boldsymbol{\nu}~\to~\max\{\mathbf{0},\boldsymbol{\nu}+\lambda \mathbf{S}_\rho^{(j_0)}\}
\end{equation}
where $\lambda>0$ is a constant and $\mathbf{S}_\rho^{(k)}$ is the $k$-th row of matrix $\mathbf{S}_\rho$; go to 2 and iterate.
\end{itemize}
This is a classical relaxation procedure. Intuitively, a violated constraint ($y_j<0$) signals that the consumption of metabolite $j$ exceeds its production. The flux of reactions where $j$ participates is then modified so as to increase the production flux and decrease the consumption flux. This procedure converges to a solution for each $\rho<1$ \cite{Martino:2007zt} so that the solutions in the limit $\rho\to 1$, where the original problem is recovered, can be obtained by increasing $\rho$ recursively to approach $1$ with the desired precision and extrapolating. Notice that the form of the update step (\ref{up}) guarantees that the bounds of variability $0\leq \nu_i<\infty$ are satisfied.

Disposing of large sets of solutions allows to evaluate many quantities of interest like the statistics of production profiles, marginal flux distributions, flux-flux correlations and, of course, distributions of ``macroscopic'' observables like, in our case, the OGI or the CMR (which are defined through simple functions of the individual fluxes). This approach has been applied in different contexts to quantify the metabolic capabilities of cells \cite{Martelli:2009jl,granata,Kyoto10}. Following the above procedure, different solutions can be generated by re-initializing the algorithm from different flux vectors. A crucial question concerns the statistics of the solution space sampling that can thus be obtained. This problem has been faced in \cite{ploscb}. In brief, if one fixes the probability distribution from which initial conditions are generated (`priors' for short), repeated iteration of the above scheme provides a set of solutions that minimize the {\it average} Euclidean distance between the solutions and the priors. More precisely, let us assume that (for each $\rho$) initial conditions are drawn from a fixed, `trial' probability distribution $P_0(\boldsymbol{\nu})$ of flux vectors (for simplicity, one may think that $P_0(\boldsymbol{\nu})=\prod_{i=1}^N P_0^{(i)}(\nu_i)$, with prescribed distributions $P_0^{(i)}$, e.g. uniform over a given interval: in this case each initial $v_i$ is selected randomly and independently from its trial distribution $P_0^{(i)}$). Then the solutions $\boldsymbol{\nu}^\star$ obtained by the above method are such that the quantity
\begin{equation}\label{d}
d^2=\avg{\sum_{i=1}^N(\nu_i^\star-\nu_i)^2}
\end{equation}
is minimized (the average being taken over $P_0$). In other words, one obtains a set of solutions that are as close as possible to the priors used to generate them and the statistical significance of the corresponding distributions of fluxes (or other relevant macroscopic observables like the OGI etc.) can be interpreted in this light. In essence, multiple (random) initializations of the above algorithm deform the uncorrelated trial distributions $P_0^{(i)}$ to generate a set of correlated probability distributions for the $\nu_i$'s, the correlation being driven by the form of the reinforcement term. Note that, quite importantly, the resulting $\nu_i$'s can exceed the initial bounds defined by $P_0$.

Now it is clear that the solution space picture one obtains can depend strongly on the choice of the priors. On one hand, disposing of sufficient empirical information about individual fluxes one can inject it into the prior (i.e. into $P_0^{(i)}$, e.g. by assuming that such distributions are uniform and centered around the empirical value) to evaluate the extent to which the solution space is constrained by sampling configurations ``close'' (on average) to such a prior. However, in the case we consider, such an information is not available in the amount and precision that would be needed. In situations like these it is reasonable to think that priors should inject into the problem as little information as possible so as to obtain unbiased information on the solution space (besides the global constraint imposed by te minimization of (\ref{d})). This approach has the advantage of providing a portrait of the solution space that is minimally constrained by the prior, in the sense that the emergent features are not due to the external biochemical information employed but, strictly speaking, to the topology of the network and the functional constraints (in our case embodied only by the OGI). In this work we have followed such a prescription.

In summary, we have solved (\ref{vn2}) using the above method for the (partial) network reconstruction described in following section. For the trial functions $P_0^{(i)}$, we took uniform distributions on $[0,1]$ (as said above, the initial bounds of variability allowed for the fluxes can be exceeded by the algorithm). The only additional constraint imposed on the solution space of (\ref{vn2}) is given by the uptake of GLC to the capillary, which is fixed and tuned externally. All other fluxes are computed self-consistently. This implies that our solution space may contain unphysiologic states that we do not discount a priori. Our focus is indeed on exploring a minimally constrained solution space. Assessing the robustness of the emerging scenario against physiological data instead highlights which bounds on variables/constraints should {\it necessarily} be added to non-equilibrium steady state models in order to reproduce a realistic phenomenology.

~

~

\section{Network reconstruction: details}

We reconstruct the cerebral metabolism via a compartmentalized model divided into blood capillary ($c$), extracellular space ($e$), neuron ($n$) and astrocyte ($a$). The last two compartments are then divided into cytosol ($nc$ and $ac$ respectively) and mitochondria ($nm$ and $am$), while we also consider vesicles ($nv$) in synapses, from where the neurotransmitter glutamate is released to the extracellular medium. Considering the reversibility of reactions the network altogether encompasses 139 chemical reactions, reported in table \ref{reactions}, which process 108 chemical species, listed in table \ref{metabolites}. Reaction names are determined by the catalyzing enzymes or, in the case of transport processes, by the chemical compound itself. The compartments involved by the chemical processes are also specified, so that, for example, ${\rm \nu HK}(n)$ denotes the reaction catalyzed by the hexokinase in neurons, while ${\rm \nu O_2} (c \to n) $ indicates an oxygen transport from the capillary to the neuron. In case of reversible reactions, the forward and reverse processes are separated into two different reactions, labeled with $f$ and $r$ respectively.

Since we are mainly interested in the carbohydrate metabolism and related energetic production, we specifically consider pathways involved in those processes plus some other features peculiar of cerebral cells, like for instance creatine synthesis and metabolism. Apart from oxygen, lactate and glucose transport from/to the capillary vessel and within the system, we have considered 14 main paths, which allow the system to self-sustain itself. We included glycolysis, pentose phosphate pathway, Tricarboxylic acid cycle (TCA), oxidative phosphorylation, which are all inherent to carbohydrate metabolism. To allow detoxification of species produced by the latter we then included the Glutathione-ascorbate (GSH/ASC) cycle, while transport of reducing equivalents (NADH) is performed through the Malate Aspartate shuttle, in turn coupled with the pentose phosphate pathway and the NAD/NADP interconversion. The coupling between functionality and metabolism is assured through neurotransmission and transmitter recycling, which together with the ionic movements provides a metabolic interconnection between neurons and astrocytes.

\clearpage

\begin{longtable*}{c}
\caption{\label{reactions} List of reactions.}\\
\begin{tabular}{|p{0.05\columnwidth}p{0.19\columnwidth}|p{0.65 \columnwidth}|}
\hline
{\em ~}&{\em Abbreviation} & {\em  Chemical Reaction} \\
\hline \hline
\end{tabular}\\
\endhead \vspace{0.2cm}
~\\{\bf GLC, O$_2$ and LAC Transport} \vspace{0.2cm}\\~\\
\begin{tabular}{|p{0.05\columnwidth}p{0.19\columnwidth}|p{0.65 \columnwidth}|}
\hline
{\footnotesize 1} & ${\rm \nu GLC}(c)$ & ${\rm \to GLC_c }$\\
{\footnotesize 2} & ${\rm \nu GLC}(c\to a)$ & ${\rm GLC_c \to GLC_a }$\\
{\footnotesize 3} & ${\rm \nu GLC}(c \to e)$ & ${\rm GLC_c \to GLC_e }$\\
{\footnotesize 4} & ${\rm \nu GLC}( e \to a)$ & ${\rm GLC_e \to GLC_a }$\\
{\footnotesize 5} & ${\rm \nu GLC}(e \to n)$ & ${\rm GLC_e \to GLC_n }$\\
{\footnotesize 6} & ${\rm \nu O_2} (c) $ & ${\rm \to {O_2}_c }$\\
{\footnotesize 7} & ${\rm \nu O_2}(c\to a)$ & ${\rm {O_2}_c \to {O_2}_a }$\\
{\footnotesize 8} & ${\rm \nu O_2} (c \to n) $ & ${\rm {O_2}_c \to {O_2}_n }$\\
{\footnotesize 9} & ${\rm \nu LAC} (c)$ & ${\rm LAC_c \to }$\\
{\footnotesize 10} & ${\rm \nu LAC}(e \to c)$ & ${\rm LAC_e \to LAC_c }$\\
{\footnotesize 11} & ${\rm \nu LAC}(a \to c)$ & ${\rm LAC_a \to LAC_c}$\\
{\footnotesize 12} & ${\rm \nu LAC}(e \to a)$ & ${\rm LAC_e \to LAC_a }$\\
{\footnotesize 13} & ${\rm \nu LAC}(a \to e)$ & ${\rm LAC_a \to LAC_e }$\\
{\footnotesize 14} & ${\rm \nu LAC}( e \to n)$ & ${\rm LAC_e \to LAC_n }$\\
{\footnotesize 15} & ${\rm \nu LAC}(n \to e)$ & ${\rm LAC_n \to LAC_e }$\\
\hline
\end{tabular}\\ \vspace{0.2cm}
~\\{\bf Creatine \& Adenylate Kinase Buffers} \vspace{0.2cm}\\~\\
\begin{tabular}{|p{0.05\columnwidth}p{0.19\columnwidth}|p{0.65 \columnwidth}|}\hline
{\footnotesize 16} & ${\rm \nu CK^f}(n)$ & ${\rm ATP_n + Cr_n \to ADP_n + PCr_n }$\\
{\footnotesize 17} & ${\rm \nu CK^r}(n)$ & ${\rm ADP_n + PCr_n \to ATP_n + Cr_n }$\\
{\footnotesize 18} & ${\rm \nu CK^f}(a)$ & ${\rm ATP_a + Cr_a \to ADP_a + PCr_a }$\\
{\footnotesize 19} & ${\rm \nu CK^r}(a)$ & ${\rm ADP_a + PCr_a \to ATP_a + Cr_a }$\\
{\footnotesize 20} & ${\rm \nu AK^f}(n)$ & ${\rm 2~ ADP_n \to AMP_n + ATP_n }$\\
{\footnotesize 21} & ${\rm \nu AK^r}(n)$ & ${\rm AMP_n + ATP_n \to 2~ ADP_n }$\\
{\footnotesize 22} & ${\rm \nu AK^f}(a)$ & ${\rm 2~ ADP_a \to AMP_a + ATP_a }$\\
{\footnotesize 23} & ${\rm \nu AK^r}(a)$ & ${\rm AMP_a + ATP_a \to 2~ ADP_a }$\\
\hline
\end{tabular}\\ \vspace{0.2cm}
~\\{\bf Pentose Phosphate Pathway} \vspace{0.2cm}\\~\\
\begin{tabular}{|p{0.05\columnwidth}p{0.19\columnwidth}|p{0.65 \columnwidth}|}\hline
{\footnotesize 24} & ${\rm \nu PPP}(n)$ & ${\rm 3~ G6P_n + 6~ NADP_n \to 2~ F6P_n + GAP_n + 6~ NADPH_n }$\\
{\footnotesize 25} & ${\rm \nu PPP}(a)$ & ${\rm 3~ G6P_a + 6~ NADP_a \to 2~ F6P_a + GAP_a + 6~ NADPH_a }$\\
\hline
\end{tabular}\\ \vspace{0.2cm}
~\\{\bf Ionic movements} \vspace{0.2cm}\\~\\
\begin{tabular}{|p{0.05\columnwidth}p{0.19\columnwidth}|p{0.65 \columnwidth}|}\hline
{\footnotesize 26} & ${\rm \nu NAK}(n)$ & ${\rm ATP_n + 2~ K_e + 3~ Na_n \to ADP_n + 2~ K_n + 3~ Na_e }$\\
{\footnotesize 27} & ${\rm \nu NAK}(a)$ & ${\rm ATP_a + 2~ K_e + 3~ Na_a \to ADP_a + 2~ K_a + 3~ Na_e }$\\
{\footnotesize 28} & ${\rm \nu NKCC}(a)$ & ${\rm K_e + Na_e \to K_a + Na_a }$\\
{\footnotesize 29} & ${\rm \nu Na}(n)$& ${\rm Na_e \to Na_n }$\\
{\footnotesize 30} & ${\rm \nu K}(n)$& ${\rm K_n \to K_e }$\\
\hline
\end{tabular}\\ \newpage \vspace{0.2cm}
~\\{\bf GSH/ASC Cycle (metabolism and transport)} \vspace{0.2cm}\\~\\
\begin{tabular}{|p{0.05\columnwidth}p{0.19\columnwidth}|p{0.65 \columnwidth}|}\hline
{\footnotesize 31} & ${\rm \nu GR^1(n)}$ & ${\rm GSSG_n + NADPH_n \to GSH_n + NADP_n }$\\
{\footnotesize 32} & ${\rm \nu GR^2}(n)$ & ${\rm GSSG_n + NADPH_{nm} \to GSH_n + NADP_{nm} }$\\
{\footnotesize 33} & ${\rm \nu DHAR}(n)$ & ${\rm DHA_n + GSH_n \to ASC_n + GSSG_n }$\\
{\footnotesize 34} & ${\rm \nu APX}(n)$ & ${\rm ASC_n + ROS_n \to DHA_n }$\\
{\footnotesize 35} & ${\rm \nu GR^1}(a)$ & ${\rm GSSG_a + NADPH_a \to GSH_a + NADP_a }$\\
{\footnotesize 36} & ${\rm \nu GR^2}(a)$ & ${\rm GSSG_a + NADPH_{am} \to GSH_a + NADP_{am} }$\\
{\footnotesize 37} & ${\rm \nu DHAR}(a)$ & ${\rm DHA_a + GSH_a \to ASC_a + GSSG_a }$\\
{\footnotesize 38} & ${\rm \nu APX}(a)$ & ${\rm ASC_a + ROS_a \to DHA_a }$\\
{\footnotesize 39} & ${\rm \nu GSH}(a)$ & ${\rm \to GSH_a }$\\
{\footnotesize 40} & ${\rm \nu GSH} (a \to e) $ & ${\rm GSH_a \to GSH_e }$\\
{\footnotesize 41} & ${\rm \nu GSH}(e \to n)$ & ${\rm GSH_e \to GSH_n }$\\
{\footnotesize 42} & ${\rm \nu DHA} (n \to e)$ & ${\rm DHA_n \to DHA_e }$\\
{\footnotesize 43} & ${\rm \nu DHA}(e \to n)$ & ${\rm DHA_e \to DHA_n }$\\
{\footnotesize 44} & ${\rm \nu DHA}( e \to a)$ & ${\rm DHA_e \to DHA_a }$\\
{\footnotesize 45} & ${\rm \nu DHA}(a \to e)$ & ${\rm DHA_a \to DHA_e }$\\
{\footnotesize 46} & ${\rm \nu ASC}(a\to e)$ & ${\rm ASC_a \to ASC_e }$\\
{\footnotesize 47} & ${\rm \nu ASC}( e \to a)$ & ${\rm ASC_e \to ASC_a }$\\
{\footnotesize 48} & ${\rm \nu ASC}(e \to n)$ & ${\rm ASC_e + 2~ Na_e \to ASC_n + 2~ Na_n }$\\
\hline
\end{tabular}\\ \vspace{0.2cm}
~\\{\bf Neurotransmission and GLU/GLN Cycle} \vspace{0.2cm}\\~\\
\begin{tabular}{|p{0.05\columnwidth}p{0.19\columnwidth}|p{0.65 \columnwidth}|}\hline
{\footnotesize 49} & ${\rm \nu GLU}( e \to a)$ & ${\rm GLU_e + K_a + 3~ Na_e \to GLU_{ac} + K_e + 3~ Na_a }$\\
{\footnotesize 50} & ${\rm \nu GS}(a)$ & ${\rm ATP_a + GLU_{ac} \to ADP_a + GLN_a }$\\
{\footnotesize 51} & ${\rm \nu GLN}(a \to n)$ & ${\rm GLN_a \to GLN_n }$\\
{\footnotesize 52} & ${\rm \nu PAG}(n)$ & ${\rm GLN_n \to GLU_{nc} }$\\
{\footnotesize 53} & ${\rm \nu GLU}(n)$ & ${\rm ATP_n + GLU_{nc} \to ADP_n + GLUnv }$\\
{\footnotesize 54} & ${\rm \nu NT}(n \to e)$ & ${\rm GLUnv \to GLU_e }$\\
\hline
\end{tabular}\\  \vspace{0.2cm}
~\\{\bf Glycolysis and Glycogenolysis} \vspace{0.2cm}\\~\\
\begin{tabular}{|p{0.05\columnwidth}p{0.19\columnwidth}|p{0.65 \columnwidth}|}\hline
{\footnotesize 55} & ${\rm \nu HK}(n)$ & ${\rm ATP_n + GLC_n \to ADP_n + G6P_n }$\\
{\footnotesize 56} & ${\rm \nu PFK}(n)$ & ${\rm ATP_n + G6P_n \to ADP_n + 2~ GAP_n }$\\
{\footnotesize 57} & ${\rm \nu GAPDH^f}(n)$ & ${\rm GAP_n + NAD_{nc} \to BPG_n + NADH_{nc} }$\\
{\footnotesize 58} & ${\rm \nu GAPDH^r}(n)$ & ${\rm BPG_n + NADH_{nc} \to GAP_n + NAD_{nc} }$\\
{\footnotesize 59} & ${\rm \nu PGK^f}(n)$ & ${\rm ADP_n + BPG_n \to ATP_n + PEP_n }$\\
{\footnotesize 60} & ${\rm \nu PGK^r}(n)$ & ${\rm ATP_n + PEP_n \to ADP_n + BPG_n }$\\
{\footnotesize 61} & ${\rm \nu PK}(n)$ & ${\rm ADP_n + PEP_n \to ATP_n + PYR_{nc} }$\\
{\footnotesize 62} & ${\rm \nu LDH^f}(n)$ & ${\rm NADH_{nc} + PYR_{nc} \to LAC_n + NAD_{nc} }$\\
{\footnotesize 63} & ${\rm \nu LDH^r}(n)$ & ${\rm LAC_n + NAD_{nc} \to NADH_{nc} + PYR_{nc} }$\\
{\footnotesize 64} & ${\rm \nu HK}(a)$ & ${\rm ATP_a + GLC_a \to ADP_a + G6P_a }$\\
{\footnotesize 65} & ${\rm \nu PFK}(a)$ & ${\rm ATP_a + G6P_a \to ADP_a + 2~ GAP_a }$\\
{\footnotesize 66} & ${\rm \nu GAPDH^f}(a)$ & ${\rm GAP_a + NAD_{ac} \to BPG_a + NADH_{ac} }$\\
{\footnotesize 67} & ${\rm \nu GAPDH^r}(a)$ & ${\rm BPG_a + NADH_{ac} \to GAP_a + NAD_{ac} }$\\
{\footnotesize 68} & ${\rm \nu PGK^f}(a)$ & ${\rm ADP_a + BPG_a \to ATP_a + PEP_a }$\\
{\footnotesize 69} & ${\rm \nu PGK^r}(a)$ & ${\rm ATP_a + PEP_a \to ADP_a + BPG_a }$\\
{\footnotesize 70} & ${\rm \nu PK}(a)$ & ${\rm ADP_a + PEP_a \to ATP_a + PYR_{ac} }$\\
{\footnotesize 71} & ${\rm \nu LDH^f}(a)$ & ${\rm NADH_{ac} + PYR_{ac} \to LAC_a + NAD_{ac} }$\\
{\footnotesize 72} & ${\rm \nu LDH^r}(a)$ & ${\rm LAC_a + NAD_{ac} \to NADH_{ac} + PYR_{ac} }$\\
\hline
\end{tabular}\\ \newpage \vspace{0.2cm}
~\\{\bf Pyruvate Shuttling to Mitochondria} \vspace{0.2cm}\\~\\
\begin{tabular}{|p{0.05\columnwidth}p{0.19\columnwidth}|p{0.65 \columnwidth}|}\hline
{\footnotesize 73} & ${\rm \nu PYR^f}(n)$ & ${\rm PYR_{nc} \to PYR_{nm} }$\\
{\footnotesize 74} & ${\rm \nu PYR^r}(n)$ & ${\rm PYR_{nm} \to PYR_{nc} }$\\
{\footnotesize 75} & ${\rm \nu PYR^f}(a)$ & ${\rm PYR_{ac} \to PYR_{am} }$\\
{\footnotesize 76} & ${\rm \nu PYR^r}(a)$ & ${\rm PYR_{am} \to PYR_{ac} }$\\
\hline
\end{tabular}\\ \vspace{0.2cm}
~\\{\bf Oxidative Phosphorylation} \vspace{0.2cm}\\~\\
\begin{tabular}{|p{0.05\columnwidth}p{0.19\columnwidth}|p{0.65 \columnwidth}|}\hline
{\footnotesize 77} & ${\rm \nu OP}(n)$ & ${\rm 5~ ADP_n + 2~ NADH_{nm} + {O_2}_n \to 5~ ATP_n + 2~ NAD_{nm} + 0.01~ ROS_n }$\\
{\footnotesize 78} & ${\rm \nu OP}(a)$ & ${\rm 5~ ADP_a + 2~ NADH_{am} + {O_2}_a \to 5~ ATP_a + 2~ NAD_{am} + 0.01~ ROS_a }$\\
\hline
\end{tabular}\\ \vspace{0.2cm}
~\\{\bf NAD/NADP Interconversion} \vspace{0.2cm}\\~\\
\begin{tabular}{|p{0.05\columnwidth}p{0.19\columnwidth}|p{0.65 \columnwidth}|}\hline
{\footnotesize 79} & ${\rm \nu NAD}(n)$ & ${\rm NADH_{nm} + NADP_{nm} \to NADPH_{nm} + NAD_{nm} }$\\
{\footnotesize 80} & ${\rm \nu NAD}(a)$ & ${\rm NADH_{am} + NADP_{am} \to NADPH_{am} + NAD_{am} }$\\
\hline
\end{tabular}\\ \vspace{0.2cm}
~\\{\bf Anaplerosys}\vspace{0.2cm}\\~\\
\begin{tabular}{|p{0.05\columnwidth}p{0.19\columnwidth}|p{0.65 \columnwidth}|}\hline
{\footnotesize 81} & ${\rm \nu PC}(a)$ & ${\rm ATP_a + PYR_{am} \to ADP_a + OAA_{am} }$\\
{\footnotesize 82} & ${\rm \nu cME}(n)$ & ${\rm MAL_{nc} + NADP_n \to NADPH_{nc} + PYR_{nc} }$\\
{\footnotesize 83} & ${\rm \nu mME}(n)$ & ${\rm MAL_{nm} + NADP_{nm} \to NADPH_{nm} + PYR_{nm} }$\\
{\footnotesize 84} & ${\rm \nu cME}(a)$ & ${\rm MAL_{ac} + NADP_a \to NADPH_{ac} + PYR_{ac} }$\\
{\footnotesize 85} & ${\rm \nu mME}(a)$ & ${\rm MAL_{am} + NADP_{am} \to NADPH_{am} + PYR_{am} }$\\
\hline
\end{tabular}\\ \vspace{0.2cm}
~\\{\bf TCA Cycle} \vspace{0.2cm}\\~\\
\begin{tabular}{|p{0.05\columnwidth}p{0.19\columnwidth}|p{0.65 \columnwidth}|}\hline
{\footnotesize 86} & ${\rm \nu PDH}(n)$ & ${\rm CoA_n + NAD_{nm} + PYR_{nm} \to ACoA_n + NADH_{nm} }$\\
{\footnotesize 87} & ${\rm \nu CS}(n)$ & ${\rm ACoA_n + OAA_{nm} \to CIT_n + CoA_n }$\\
{\footnotesize 88} & ${\rm \nu IDH^1}(n)$ & ${\rm CIT_n + NAD_{nm} \to AKG_{nm} + NADH_{nm} }$\\
{\footnotesize 89} & ${\rm \nu IDH^2}(n)$ & ${\rm CIT_n + NADP_{nm} \to AKG_{nm} + NADPH_{nm} }$\\
{\footnotesize 90} & ${\rm \nu IDH^3}(n)$ & ${\rm CIT_n + NADP_n \to AKG_{nc} + NADPH_n }$\\
{\footnotesize 91} & ${\rm \nu AKGDH}(n)$ & ${\rm AKG_{nm} + CoA_n + NAD_{nm} \to NADH_{nm} + SCoA_n }$\\
{\footnotesize 92} & ${\rm \nu SCoATK^f}(n)$ & ${\rm ADP_n + SCoA_n \to ATP_n + SUC_n }$\\
{\footnotesize 93} & ${\rm \nu SCoATK^r}(n)$ & ${\rm ATP_n + SUC_n \to ADP_n + SCoA_n }$\\
{\footnotesize 94} & ${\rm \nu SDH^f}(n)$ & ${\rm 1.50~ ADP_n + 0.10~ {O_2}_n + 3~ SUC_n \to 1.50~ ATP_n + 3~ FUM_n }$\\
{\footnotesize 95} & ${\rm \nu SDH^r}(n)$ & ${\rm 1.50~ ATP_n + 3~ FUM_n \to 1.50~ ADP_n + 0.10~ {O_2}_n + 3~ SUC_n }$\\
{\footnotesize 96} & ${\rm \nu FUM^f}(n)$ & ${\rm FUM_n \to MAL_{nm} }$\\
{\footnotesize 97} & ${\rm \nu FUM^r}(n)$ & ${\rm MAL_{nm} \to FUM_n }$\\
{\footnotesize 98} & ${\rm \nu mMDH}(n)$ & ${\rm MAL_{nm} + NAD_{nm} \to NADH_{nm} + OAA_{nm} }$\\
{\footnotesize 99} & ${\rm \nu PDH}(a)$ & ${\rm CoA_a + NAD_{am} + PYR_{am} \to ACoA_a + NADH_{am} }$\\
{\footnotesize 100} & ${\rm \nu CS}(a)$ & ${\rm ACoA_a + OAA_{am} \to CIT_a + CoA_a }$\\
{\footnotesize 101} & ${\rm \nu IDH^1}(a)$ & ${\rm CIT_a + NAD_{am} \to AKG_{am} + NADH_{am} }$\\
{\footnotesize 102} & ${\rm \nu IDH^2}(a)$ & ${\rm CIT_a + NADP_{am} \to AKG_{am} + NADPH_{am} }$\\
{\footnotesize 103} & ${\rm \nu IDH^3}(a)$ & ${\rm CIT_a + NADP_a \to AKG_{ac} + NADPH_a }$\\
{\footnotesize 104} & ${\rm \nu AKGDH}(a)$ & ${\rm AKG_{am} + CoA_a + NAD_{am} \to NADH_{am} + SCoA_a }$\\
{\footnotesize 105} & ${\rm \nu SCoATK^f}(a)$ & ${\rm ADP_a + SCoA_a \to ATP_a + SUC_a }$\\
{\footnotesize 106} & ${\rm \nu SCoATK^r}(a)$ & ${\rm ATP_a + SUC_a \to ADP_a + SCoA_a }$\\
{\footnotesize 107} & ${\rm \nu SDH^f}(a)$ & ${\rm 1.50~ ADP_a + 0.10~ {O_2}_a + 3~ SUC_a \to 1.50~ ATP_a + 3~ FUM_a }$\\
{\footnotesize 108} & ${\rm \nu SDH^r}(a)$ & ${\rm 1.50~ ATP_a + 3~ FUM_a \to 1.50~ ADP_a + 0.10~ {O_2}_a + 3~ SUC_a }$\\
{\footnotesize 109} & ${\rm \nu FUM^f}(a)$ & ${\rm FUM_a \to MAL_{am} }$\\
{\footnotesize 110} & ${\rm \nu FUM^r}(a)$ & ${\rm MAL_{am} \to FUM_a }$\\
{\footnotesize 111} & ${\rm \nu mMDH}(a)$ & ${\rm MAL_{am} + NAD_{am} \to NADH_{am} + OAA_{am} }$\\
\hline
\end{tabular}\\ \newpage \vspace{0.2cm}
~\\{\bf Glycerol 3-P Shuttle} \vspace{0.2cm}\\~\\
\begin{tabular}{|p{0.05\columnwidth}p{0.19\columnwidth}|p{0.65 \columnwidth}|}\hline
{\footnotesize 112} & ${\rm \nu G3PS}(n)$ & ${\rm 1.50~ ADP_n + 3~ NADH_{nc} + 0.10~ {O_2}_n \to 1.50~ ATP_n + 3~ NAD_{nc} }$\\
{\footnotesize 113} & ${\rm \nu G3PS}(a)$ & ${\rm 1.50~ ADP_a + 3~ NADH_{ac} + 0.10~ {O_2}_a \to 1.50~ ATP_a + 3~ NAD_{ac} }$\\
\hline
\end{tabular}\\ \vspace{0.2cm}
~\\{\bf MAS} \vspace{0.2cm}\\~\\
\begin{tabular}{|p{0.05\columnwidth}p{0.19\columnwidth}|p{0.65 \columnwidth}|}\hline
{\footnotesize 114} & ${\rm \nu cMDH}(n)$ & ${\rm NADH_{nc} + OAA_{nc} \to MAL_{nc} + NAD_{nc} }$\\
{\footnotesize 115} & ${\rm \nu OGC}(n)$ & ${\rm AKG_{nm} + MAL_{nc} \to AKG_{nc} + MAL_{nm} }$\\
{\footnotesize 116} & ${\rm \nu AGC}(n)$ & ${\rm ASP_{nm} + GLU_{nc} \to ASP_{nc} + GLU_{nm} }$\\
{\footnotesize 117} & ${\rm \nu cMDH}(a)$ & ${\rm NADH_{ac} + OAA_{ac} \to MAL_{ac} + NAD_{ac} }$\\
{\footnotesize 118} & ${\rm \nu OGC}(a)$ & ${\rm AKG_{am} + MAL_{ac} \to AKG_{ac} + MAL_{am} }$\\
{\footnotesize 119} & ${\rm \nu AGC}(a)$ & ${\rm ASP_{am} + GLU_{ac} \to ASP_{ac} + GLU_{am} }$\\
\hline
\end{tabular}\\  \vspace{0.2cm}
~\\{\bf Aspartate Metabolism and Shuttling} \vspace{0.2cm}\\~\\
\begin{tabular}{|p{0.05\columnwidth}p{0.19\columnwidth}|p{0.65 \columnwidth}|}\hline
{\footnotesize 120} & ${\rm \nu cAAT^f}(n)$ & ${\rm AKG_{nc} + ASP_{nc} \to GLU_{nc} + OAA_{nc} }$\\
{\footnotesize 121} & ${\rm \nu cAAT^r}(n)$ & ${\rm GLU_{nc} + OAA_{nc} \to AKG_{nc} + ASP_{nc} }$\\
{\footnotesize 122} & ${\rm \nu mAAT^f}(n)$ & ${\rm AKG_{nm} + ASP_{nm} \to GLU_{nm} + OAA_{nm} }$\\
{\footnotesize 123} & ${\rm \nu mAAT^r}(n)$ & ${\rm GLU_{nm} + OAA_{nm} \to AKG_{nm} + ASP_{nm} }$\\
{\footnotesize 124} & ${\rm \nu cAAT^f}(a)$ & ${\rm AKG_{ac} + ASP_{ac} \to GLU_{ac} + OAA_{ac} }$\\
{\footnotesize 125} & ${\rm \nu cAAT^r}(a)$ & ${\rm GLU_{ac} + OAA_{ac} \to AKG_{ac} + ASP_{ac} }$\\
{\footnotesize 126} & ${\rm \nu mAAT^f}(a)$ & ${\rm AKG_{am} + ASP_{am} \to GLU_{am} + OAA_{am} }$\\
{\footnotesize 127} & ${\rm \nu mAAT^r}(a)$ & ${\rm GLU_{am} + OAA_{am} \to AKG_{am} + ASP_{am} }$\\
{\footnotesize 128} & ${\rm \nu ASP^f}(n \to a)$ & ${\rm ASP_{nc} \to ASP_{ac} }$\\
{\footnotesize 129} & ${\rm \nu ASP^r}(n \to a)$ & ${\rm ASP_{ac} \to ASP_{nc} }$\\
{\footnotesize 130} & ${\rm \nu ASP^f}(n)$ & ${\rm ASP_{nc} \to ASP_{nm} }$\\
{\footnotesize 131} & ${\rm \nu ASP^r}(n)$ & ${\rm ASP_{nm} \to ASP_{nc} }$\\
{\footnotesize 132} & ${\rm \nu ASP^f}(a)$ & ${\rm ASP_{ac} \to ASP_{am} }$\\
{\footnotesize 133} & ${\rm \nu ASP^r}(a)$ & ${\rm ASP_{am} \to ASP_{ac} }$\\
\hline
\end{tabular}\\ \vspace{0.2cm}
~\\{\bf Glutamate Metabolism}\vspace{0.2cm}\\~\\
\begin{tabular}{|p{0.05\columnwidth}p{0.19\columnwidth}|p{0.65 \columnwidth}|}\hline
{\footnotesize 134} & ${\rm \nu GDH^f}(n)$ & ${\rm GLU_{nm} + NADP_{nm} \to AKG_{nm} + NADPH_{nm} }$\\
{\footnotesize 135} & ${\rm \nu GDH^r}(n)$ & ${\rm AKG_{nm} + NADPH_{nm} \to GLU_{nm} + NADP_{nm} }$\\
{\footnotesize 136} & ${\rm \nu GDH^f}(a)$ & ${\rm GLU_{am} + NADP_{am} \to AKG_{am} + NADPH_{am} }$\\
{\footnotesize 137} & ${\rm \nu GDH^r}(a)$ & ${\rm AKG_{am} + NADPH_{am} \to GLU_{am} + NADP_{am} }$\\
\hline
\end{tabular}\\ \vspace{0.2cm}
~\\{\bf Housekeeping} \vspace{0.2cm}\\~\\
\begin{tabular}{|p{0.05\columnwidth}p{0.19\columnwidth}|p{0.65 \columnwidth}|}\hline
{\footnotesize 138} & ${\rm \nu ATP}(n)$ & ${\rm ATP_n \to ADP_n}$\\
{\footnotesize 139} & ${\rm \nu ATP}(a)$ & ${\rm ATP_a \to ADP_a }$\\
\hline
\end{tabular}
\end{longtable*}

~

~

~

~

~

~

~

~

~

~

~

~

~

~

~

~

~

~

~

~

~

~

~

~

~

~

~

~

~

~

~

~

~

~


\begin{longtable*}{c}
\caption{\label{metabolites}List of metabolites.}\\
\begin{tabular}{|p{0.025\columnwidth}p{0.1\columnwidth}p{0.3\columnwidth}|p{0.025\columnwidth}p{0.1\columnwidth}p{0.3\columnwidth}|}
\hline \hline
{\em No.}& {\em Abbr.}& {\em Name}&  {\em No.}& {\em Abbr.}& {\em Name}\\
\hline
\end{tabular}\\
\endhead \vspace{0.25cm}
~\\{\bf Neuron} \vspace{0.25cm}\\~\\
\begin{tabular}{|p{0.025\columnwidth}p{0.1\columnwidth}p{0.3\columnwidth}|p{0.025\columnwidth}p{0.1\columnwidth}p{0.3\columnwidth}|}
\hline
{\footnotesize 1} &  ${\rm ACoA_n}$  & Acetyl-CoA &        {\footnotesize 26} & ${\rm K_n}$  & Potassium\\
{\footnotesize 2} &  ${\rm ADP_n}$  & Adenosine diphosphate &         {\footnotesize 27} & ${\rm LAC_n}$  & Lactic acid \\
{\footnotesize 3} &  ${\rm AKG_{nc}}$  & $\alpha$-Ketoglutaric acid  &      {\footnotesize 28} & ${\rm MAL_{nc}}$  & Malic acid\\
{\footnotesize 4} &  ${\rm AKG_{nm}}$  & $\alpha$-Ketoglutaric acid ($m$)&      {\footnotesize 29} & ${\rm MAL_{nm}}$ & Malic acid ($m$)\\
{\footnotesize 5} &  ${\rm AMP_n}$  & Adenosine monophosphate &         {\footnotesize 30} & ${\rm NADH_{nc}}$  & N. adenine dinucleotide\\
{\footnotesize 6} &  ${\rm ASC_n}$  & Ascorbic acid &         {\footnotesize 31} & ${\rm NADH_{nm}}$  & N. adenine dinucleotide ($m$)\\
{\footnotesize 7} &  ${\rm ASP_{nc}}$  & Aspartic acid &      {\footnotesize 32} & ${\rm NAD_{nc}}$  & N. adenine dinucleotide \\
{\footnotesize 8} &  ${\rm ASP_{nm}}$  & Aspartic acid ($m$)&      {\footnotesize 33} & ${\rm NAD_{nm}}$  & N. adenine dinucleotide ($m$)\\
{\footnotesize 9} &  ${\rm ATP_n}$  & Adenosine triphosphate &         {\footnotesize 34} & ${\rm NADPH_n}$  & N. adenine dinucleotideph.\\
{\footnotesize 10} &  ${\rm BPG_n}$  & 1,3-Bisphosphoglyceric acid &        {\footnotesize 35} & ${\rm NADPH_{nc}}$  & N. adenine dinucleotideph.\\
{\footnotesize 11} &  ${\rm CIT_n}$  & Citrate &        {\footnotesize 36} & ${\rm NADPH_{nm}}$  & N. adenine dinucleotideph.($m$)\\
{\footnotesize 12} &  ${\rm CoA_n}$  & Coenzyme A &        {\footnotesize 37} & ${\rm NADP_n}$  & N. adenine dinucleotideph.\\
{\footnotesize 13} &  ${\rm Cr_n}$  & Creatine &         {\footnotesize 38} & ${\rm NADP_{nm}}$  & N. adenine dinucleotideph.($m$)\\
{\footnotesize 14} &  ${\rm DHA_n}$  & Dehydroascorbic acid &        {\footnotesize 39} & ${\rm Na_n}$  & Sodium\\
{\footnotesize 15} &  ${\rm F6P_n}$  & Fructose 6-phosphate &        {\footnotesize 40} & ${\rm O2_n}$  & Oxygen\\
{\footnotesize 16} &  ${\rm FUM_n}$  & Fumaric acid &        {\footnotesize 41} & ${\rm OAA_{nc}}$  & Oxaloacetic acid\\
{\footnotesize 17} &  ${\rm G6P_n}$  & Glucose 6-phosphate &        {\footnotesize 42} & ${\rm OAA_{nm}}$  & Oxaloacetic acid\\
{\footnotesize 18} &  ${\rm GAP_n}$  & Glyceraldehyde  3-phosphate &        {\footnotesize 43} & ${\rm PCr_n}$  & Phosphocreatine\\
{\footnotesize 19} &  ${\rm GLC_n}$  & Glucose &        {\footnotesize 44} & ${\rm PEP_n}$  & Phosphoenolpyruvic acid\\
{\footnotesize 20} &  ${\rm GLN_n}$  & Glutamine &        {\footnotesize 45} & ${\rm PYR_{nc}}$  & Pyruvic acid\\
{\footnotesize 21} &  ${\rm GLU_{nc}}$  & Glutamate &     {\footnotesize 46} & ${\rm PYR_{nm}}$  & Pyruvic acid ($m$)\\
{\footnotesize 22} &  ${\rm GLU_{nm}}$  & Glutamate ($m$)&     {\footnotesize 47} & ${\rm ROS_n}$  & Reactive oxygen species\\
{\footnotesize 23} &  ${\rm GLU_{nv}}$  & Glutamate (vescicle) &     {\footnotesize 48} & ${\rm SCoA_n}$  & Succinyl-CoA\\
{\footnotesize 24} &  ${\rm GSH_n}$  & Glutathione &        {\footnotesize 49} & ${\rm SUC_n}$  & Succinic acid\\
{\footnotesize 25} &  ${\rm GSSG_n}$  & Glutathione disulfide & && \\
\hline
\end{tabular}\\ \newpage \vspace{0.25cm}
~\\{\bf Astrocyte} \vspace{0.25cm} \\~\\
\begin{tabular}{|p{0.025\columnwidth}p{0.1\columnwidth}p{0.3\columnwidth}|p{0.025\columnwidth}p{0.1\columnwidth}p{0.3\columnwidth}|}
\hline
{\footnotesize 50} &  ${\rm ACoA_a}$  & Acetyl-CoA &       {\footnotesize 74} & ${\rm K_a}$  & Potassium \\
{\footnotesize 51} &  ${\rm ADP_a}$  & denosine diphosphate &        {\footnotesize 75} & ${\rm LAC_a}$  & Lactic acid \\
{\footnotesize 52} &  ${\rm AKG_{ac}}$  & $\alpha$-Ketoglutaric acid &     {\footnotesize 76} & ${\rm MAL_{ac}}$  & Malic acid \\
{\footnotesize 53} &  ${\rm AKG_{am}}$  & $\alpha$-Ketoglutaric acid ($m$)&     {\footnotesize 77} & ${\rm MAL_{am}}$  & Malic acid ($m$)\\
{\footnotesize 54} &  ${\rm AMP_a}$  & Adenosine monophosphate &        {\footnotesize 78} & ${\rm Na_a}$  & Sodium \\
{\footnotesize 55} &  ${\rm ASC_a}$  & Ascorbic acid &        {\footnotesize 79} & ${\rm NAD_{ac}}$  & N. adenine dinucleotide \\
{\footnotesize 56} &  ${\rm ASP_{ac}}$  & Aspartic acid &     {\footnotesize 80} & ${\rm NAD_{am}}$  & N. adenine dinucleotide ($m$)\\
{\footnotesize 57} &  ${\rm ASP_{am}}$  & Aspartic acid ($m$)&     {\footnotesize 81} & ${\rm NADH_{ac}}$  & N. adenine dinucleotide \\
{\footnotesize 58} &  ${\rm ATP_a}$  & Adenosine triphosphate &        {\footnotesize 82} & ${\rm NADH_{am}}$  & N. adenine dinucleotide ($m$)\\
{\footnotesize 59} &  ${\rm BPG_a}$  & 1,3-Bisphosphoglyceric acid &        {\footnotesize 83} & ${\rm NADP_a}$  & N. adenine dinucleotideph.\\
{\footnotesize 60} &  ${\rm CIT_a}$  & Citrate &        {\footnotesize 84} & ${\rm NADP_{am}}$  & N. adenine dinucleotideph. ($m$)\\
{\footnotesize 61} &  ${\rm CoA_a}$  & Coenzyme A &        {\footnotesize 85} & ${\rm NADPH_a}$  & N. adenine dinucleotideph.\\
{\footnotesize 62} &  ${\rm Cr_a}$  & Creatine &         {\footnotesize 86} & ${\rm NADPH_{ac}}$  & N. adenine dinucleotideph.\\
{\footnotesize 63} &  ${\rm DHA_a}$  & Dehydroascorbic acid &        {\footnotesize 87} & ${\rm NADPH_{am}}$  & N. adenine dinucleotideph. ($m$)\\
{\footnotesize 64} &  ${\rm F6P_a}$  & Fructose 6-phosphate &        {\footnotesize 88} & ${\rm O2_a}$  & Oxygen \\
{\footnotesize 65} &  ${\rm FUM_a}$  & Fumaric acid &        {\footnotesize 89} & ${\rm OAA_{ac}}$  & Oxaloacetic acid \\
{\footnotesize 66} &  ${\rm G6P_a}$  & Glucose 6-phosphate &        {\footnotesize 90} & ${\rm OAA_{am}}$  & Oxaloacetic acid ($m$)\\
{\footnotesize 67} &  ${\rm GAP_a}$  & Glyceraldehyde 3-phosphate &        {\footnotesize 91} & ${\rm PCr_a}$  & Phosphocreatine \\
{\footnotesize 68} &  ${\rm GLC_a}$  & Glucose &        {\footnotesize 92} & ${\rm PEP_a}$  & Phosphoenolpyruvic acid \\
{\footnotesize 69} &  ${\rm GLN_a}$  & Glutamine &        {\footnotesize 93} & ${\rm PYR_{ac}}$  & Pyruvic acid \\
{\footnotesize 70} &  ${\rm GLU_{ac}}$  & Glutamate &     {\footnotesize 94} & ${\rm PYR_{am}}$  & Pyruvic acid ($m$)\\
{\footnotesize 71} &  ${\rm GLU_{am}}$  & Glutamate ($m$)&     {\footnotesize 95} & ${\rm ROS_a}$  & Reactive oxygen species \\
{\footnotesize 72} &  ${\rm GSH_a}$  & Glutathione &        {\footnotesize 96} & ${\rm SCoA_a}$  & Succinyl-CoA \\
{\footnotesize 73} &  ${\rm GSSG_a}$  & Glutathione disulfide &       {\footnotesize 97} & ${\rm SUC_a}$  & Succinic acid \\
\hline
\end{tabular}\\
\vspace{0.25cm}
~\\{\bf Extracellular space and blood capillary} \vspace{0.25cm} \\~\\
\begin{tabular}{|p{0.025\columnwidth}p{0.1\columnwidth}p{0.3\columnwidth}|p{0.025\columnwidth}p{0.1\columnwidth}p{0.3\columnwidth}|}
\hline
{\footnotesize 98} &  ${\rm ASC_e}$  & Ascorbic acid &        {\footnotesize 104} & ${\rm LAC_e}$  & Lactic acid \\
{\footnotesize 99} &  ${\rm DHA_e}$  & Dehydroascorbic acid &        {\footnotesize 105} & ${\rm Na_e}$  & Sodium \\
{\footnotesize 100} &  ${\rm GLC_e}$  & Glucose &       {\footnotesize 106} &  ${\rm GLC_c}$  & Glucose \\
{\footnotesize 101} &  ${\rm GLU_e}$  & Glutamate &       {\footnotesize 107} &  ${\rm LAC_c}$  & Lactic acid \\
{\footnotesize 102} & ${\rm GSH_e}$  & Glutathione & {\footnotesize 108} & ${\rm O2_c}$  & Oxygen \\
{\footnotesize 103} & ${\rm K_e}$  & Potassium & &&\\
\hline
\end{tabular}
\end{longtable*}


\begin{thebibliography}{99}
\providecommand{\url}[1]{\texttt{#1}}
\providecommand{\urlprefix}{URL }
\expandafter\ifx\csname urlstyle\endcsname\relax
  \providecommand{\doi}[1]{doi:\discretionary{}{}{}#1}\else
  \providecommand{\doi}{doi:\discretionary{}{}{}\begingroup
  \urlstyle{rm}\Url}\fi
\providecommand{\bibAnnoteFile}[1]{%
  \IfFileExists{#1}{\begin{quotation}\noindent\textsc{Key:} #1\\
  \textsc{Annotation:}\ \input{#1}\end{quotation}}{}}
\providecommand{\bibAnnote}[2]{%
  \begin{quotation}\noindent\textsc{Key:} #1\\
  \textsc{Annotation:}\ #2\end{quotation}}
\providecommand{\eprint}[2][]{\url{#2}}
%

\bibitem{Mangia2007a}
Mangia S, Giove F, Tk\'{a}c I, Logothetis NK, Olman CA, Maraviglia B, Di Salle F, Ugurbil K: {\bf Metabolic
  and hemodynamic events after changes in neuronal activity: current
  hypotheses, theoretical predictions and in vivo {NMR} experimental findings.}
\newblock {{\it J Cereb Blood Flow Metab}} 2009, {\bf 29}: 441--463.
\input{Mangia2007a}

\bibitem{Pellerin1994}
Pellerin L, Magistretti P:  {\bf Glutamate uptake into astrocytes stimulates
  aerobic glycolysis: a mechanism coupling neuronal activity to glucose
  utilization.}
\newblock {{\it Proc Natl Acad Sci USA}} 1994, {\bf 91}: 10625--10629.
\input{Pellerin1994}

\bibitem{Dienel2006}
Dienel GA, Cruz NF: {\bf Astrocyte activation in working brain: energy
  supplied by minor substrates.}
\newblock {\it Neurochem Int} 2006, {\bf 48}: 586--595.
\input{Dienel2006}

\bibitem{Lak}
Bak LK, Schousboe A, Sonnewald U and Waagepetersen HS:
{\bf Glucose is necessary to maintain neurotransmitter homeostasis during synaptic activity in cultured glutamatergic neurons.}
{\it J Cereb Blood Flow Metab} 2006, {\bf 26}: 1285--1297.

\bibitem{dien}
Dienel GA: 
{\bf Astrocytic energetics during excitatory neurotransmission: What are contributions of glutamate oxidation and glycolysis?} 
{\it Neurochem Int} 2013, {\bf 63}: 244--258.

\bibitem{merg}
Mergenthaler P, Lindauer U, Dienel GA, Meisel A:
{\bf Sugar for the brain: the role of glucose in physiological and pathological brain function} (2013).
{\it Trends Neurosci} 2013, in press.

\bibitem{Aubert2005}
Aubert A, Costalat R:
{\bf Interaction between astrocytes and neurons studied
  using a mathematical model of compartmentalized energy metabolism.}
\newblock {\it J Cereb Blood Flow Metab} 2005,  {\bf 25}: 1476--1490.
\input{Aubert2005}

\bibitem{Aubert2007}
Aubert A, Pellerin L, Magistretti PJ, Costalat R:
{\bf A coherent
  neurobiological framework for functional neuroimaging provided by a model
  integrating compartmentalized energy metabolism.}
\newblock {\it Proc Natl Acad Sci USA} 2007, {\bf 104}: 4188--4193.
\input{Aubert2007}

\bibitem{Simpson2007}
Simpson IA, Carruthers A, Vannucci SJ:
{\bf Supply and demand in cerebral
  energy metabolism: the role of nutrient transporters.}
\newblock {\it J Cereb Blood Flow Metab} 2007, {\bf 27}: 1766--1791.
\input{Simpson2007}

\bibitem{Mangia2009b}
Mangia S, Simpson IA, Vannucci SJ, Carruthers A:
{\bf The in vivo
  neuron-to-astrocyte lactate shuttle in human brain: evidence from modeling of
  measured lactate levels during visual stimulation.}
\newblock {\it J Neurochem} 2009, {\bf 109} Suppl 1: 55--62.
\input{Mangia2009b}

\bibitem{DiNuzzo2010c}
{Di Nuzzo} M, Mangia S, Maraviglia B, Giove F:
{\bf Changes in glucose uptake
  rather than lactate shuttle take center stage in subserving neuroenergetics:
  evidence from mathematical modeling.}
\newblock {\it J Cereb Blood Flow Metab} 2010,  {\bf 30}: 586--602.
\input{DiNuzzo2010c}

\bibitem{Somersalo2012}
Somersalo E, Cheng Y, Calvetti D:
{\bf The Metabolism of Neurons and Astrocytes Through Mathematical
Models.}
\newblock {\it Annals of Biomedical Engineering} 2012,  {\bf 40}: 2328--2344.
\input{Somersalo2012}

\bibitem{Calvetti2011}
Calvetti D, Somersalo E:
{\bf Dynamic activation model for a glutamatergic
  neurovascular unit.}
\newblock {\it J Theor Biol} 2011, {\bf 274}: 12--29.
\input{Calvetti2011}

\bibitem{Jolivet2009}
Jolivet R, Magistretti PJ, Weber B:
{\bf Deciphering neuron-glia
  compartmentalization in cortical energy metabolism.}
\newblock {\it Front Neuroenergetics} 2009, {\bf 1}: 4.
\input{Jolivet2009}

\bibitem{Occhipinti2009}
Occhipinti R, Somersalo E, Calvetti D:
{\bf Astrocytes as the glucose shunt
  for glutamatergic neurons at high activity: an in silico study.}
\newblock {\it J Neurophysiol} 2009, {\bf 101}: 2528--2538.
\input{Occhipinti2009}

\bibitem{Dienel2012a}
Dienel GA:
{\bf Brain lactate metabolism: the discoveries and the controversies.}
\newblock {\it J Cereb Blood Flow Metab} 2012,  {\bf 32}: 1107-38.
\input{Dienel2012a}

\bibitem{Jolivet2010}
Jolivet R, Allaman I, Pellerin L, Magistretti PJ, Weber B:
{\bf Comment on
  recent modeling studies of astrocyte-neuron metabolic interactions.}
\newblock {\it J Cereb Blood Flow Metab} 2010,  {\bf 30}: 1982--1986.
\input{Jolivet2010}

\bibitem{casom2012}
Calvetti D, Somersalo E:
{\bf M\'{e}nage \`{a} trois: the role of neurotransmitters in the energy metabolism of astrocytes, glutamatergic, and GABAergic neurons.}
\newblock {\it J Cereb Blood Flow Metab} 2012, {\bf 32}: 1472--1783.


\bibitem{Mangia2011}
Mangia S, {Di Nuzzo} M, Giove F, Carruthers A, Simpson IA, Vannucci SJ:
  {\bf Response to '{C}omment on recent modeling studies of astrocyte-neuron
  metabolic interactions': much ado about nothing.}
\newblock {\it J Cereb Blood Flow Metab} 2011, {\bf 31}: 1346--1353.
\input{Mangia2011}

\bibitem{Dienel2012b}
Dienel GA:
{\bf Fueling and imaging brain activation.}
\newblock {\it ASN Neuro ASN Neuro} 2012, {\bf 4}: e00093.
\input{Dienel2012b}


\bibitem{Martelli2009}
Martelli C, {De Martino} A, Marinari E, Marsili M, Castillo IP:
  {\bf Identifying essential genes in escherichia coli from a metabolic optimization
  principle.}
\newblock {\it Proc Natl Acad Sci USA} 2009, {\bf 106}: 2607--2611.
\input{Martelli2009}

\bibitem{ploscb}
De Martino D, Figliuzzi M, De Martino A, Marinari E:
{\bf A Scalable Algorithm to Explore the Gibbs Energy Landscape of Genome-Scale Metabolic Networks.}
{\it PLoS Comput Biol} 2012,  {\bf 8}: e1002562.

\bibitem{Tunahan2007}
Tunahan {\c{C}}, Selma A, Hale S, Ata A, Kutlu {\"U}:
{\bf Reconstruction and
  flux analysis of coupling between metabolic pathways of astrocytes and
  neurons: application to cerebral hypoxia}.
\newblock {\it Theor Biol Med Model} 2007, {\bf 4}: 48.
\input{Tunahan2007}

\bibitem{Schellenberger-RBC} Price ND, J Schellenberger, B\O~ Palsson:
{\bf Uniform
  sampling of steady-state flux spaces: means to design experiments
  and to interpret enzymopathies.}
  {\it Biophys J} 2004,  {\bf 87}: 2172--2186.

\bibitem{Hyder1998}
{Hyder F, Shulman RG, Rothman DL}:
{\bf A model for the regulation of
  cerebral oxygen delivery}.
\newblock {\it J Appl Physiol} 1998,  {\bf 85}: {554-564}.
\input{Hyder1998}

\bibitem{Gjedde1997}
Gjedde A:
{\bf The relation between brain function and cerebral blood flow and
  metabolism.}
\newblock In {\it Cerebrovascular Disease}. Edited by  Batjer HH, Philadelphia:
  Lippincott--Raven 1997: 23--40
\input{Gjedde1997}

\bibitem{Lovatt2007}
Lovatt D, Sonnewald U, Waagepetersen HS, Schousboe A, He W, Lin JH, Han X, Takano T, Wang S, Sim FJ, Goldman SA, Nedergaard M:
{\bf The transcriptome and metabolic gene signature of protoplasmic astrocytes in the
  adult murine cortex.}
\newblock {\it J Neurosci} 2007, {\bf 27}: 12255--12266.
\input{Lovatt2007}

\bibitem{Hertz2011}
Hertz L:
{\bf Brain glutamine synthesis requires neuronal aspartate: a
  commentary.}
\newblock {\it J Cereb Blood Flow Metab} 2011,  {\bf 31}: 384--387.
\input{Hertz2011}

\bibitem{Dringen1999}
Dringen R, Pfeiffer B, Hamprecht B:
{\bf Synthesis of the antioxidant
  glutathione in neurons: supply by astrocytes of cysgly as precursor for
  neuronal glutathione.}
\newblock {\it J Neurosci} 1999, {\bf 19}: 562--569.
\input{Dringen1999}

\bibitem{Gegg2003}
Gegg ME, Beltran B, Salas-Pino S, Bolanos JP, Clark JB, Moncada S, Heales SJ:
  {\bf Differential effect of nitric oxide on glutathione metabolism and
  mitochondrial function in astrocytes and neurones: implications for
  neuroprotection/neurodegeneration?}
\newblock {\it J Neurochem} 2003, {\bf 86}: 228--237.
\input{Gegg2003}

\bibitem{Yu1983}
Yu AC, Drejer J, Hertz L, Schousboe A:
{\bf Pyruvate carboxylase activity in
  primary cultures of astrocytes and neurons.}
\newblock {\it J Neurochem} 1983, {\bf 41}: 1484--1487.
\input{Yu1983}

\bibitem{Hassel2000}
Hassel B, Br\aa the A:
{\bf Neuronal pyruvate carboxylation supports formation
  of transmitter glutamate.}
\newblock {\it J Neurosci} 2000, {\bf 20}: 1342--1347.
\input{Hassel2000}

\bibitem{Ames2000}
Ames A:
{\bf {CNS} energy metabolism as related to function.}
\newblock {\it Brain Res Brain Res Rev} 2000, {\bf 34}: 42--68.
\input{Ames2000}

\bibitem{Danbolt2001}
Danbolt NC
{\bf Glutamate uptake.}
\newblock {\it  Prog Neurobiol} 2001, {\bf 65}: 1--105.


\bibitem{Huang2004}
Huang YH, Bergles DE
{\bf Glutamate transporters bring competition to the synapse.}
\newblock {\it  Curr Opin Neurobiol} 2004, {\bf 14}: 346--352.

\bibitem{mangiancr}
Mangia S, Giove F, Di Nuzzo M
{\bf Metabolic Pathways and Activity-Dependent Modulation of Glutamate Concentration in the Human Brain.}
{\it Neurochem Res} 2012, {\bf 37}: 2554--2561.

\bibitem{Waagepetersen2000}
Waagepetersen HS, Sonnewald U, Larsson OM, Schousboe A:
{\bf A possible role
  of alanine for ammonia transfer between astrocytes and glutamatergic neurons.}
\newblock {\it J Neurochem} 2000, {\bf 75}: 471--479.
\input{Waagepetersen2000}

\bibitem{Bak2005}
Bak LK, Sickmann HM, Schousboe A, Waagepetersen HS:
{\bf Activity of the
  lactate-alanine shuttle is independent of glutamate-glutamine cycle activity
  in cerebellar neuronal-astrocytic cultures.}
\newblock {\it J Neurosci Res} 2005,  {\bf 79}: 88--96.
\input{Bak2005}


\bibitem{Mangia2007}
Mangia S, Tk\'{a}c I, Gruetter R, {Van de Moortele} PF, Maraviglia B, Ugurbil K:
  {\bf {S}ustained neuronal activation raises oxidative metabolism to a new
  steady-state level: evidence from (1){H} {NMR} spectroscopy in the human
  visual cortex.}
\newblock {\it J Cereb Blood Flow Metab}   2007,  {\bf 27}: 1055--1063.
\input{Mangia2007}

\bibitem{Bandettini1997a}
Bandettini PA, Kwong KK, Davis TL, Tootell RB, Wong EC, Fox PT, Belliveau JW, Weisskoff RM, Rosen BR:
  {\bf Characterization of cerebral blood oxygenation and flow changes during
  prolonged brain activation.}
\newblock {\it Hum Brain Mapp} 1997, {\bf 5}: 93--109.
\input{Bandettini1997a}

\bibitem{Logothetis2001}
Logothetis NK, Pauls J, Augath M, Trinath T, Oeltermann A:
  {\bf Neurophysiological investigation of the basis of the {fMRI} signal.}
\newblock {\it Nature} 2001, {\bf 412}: 150--157.
\input{Logothetis2001}

\bibitem{palsson0}
Palsson B. \O. {\it Systems biology: Properties of reconstructed networks}. Cambridge (UK): Cambridge University Press 2006.

\bibitem{recent}
Schellenberger J, Palsson B:
{\bf Use of Randomized Sampling for Analysis of Metabolic Networks.}
{\it J. Biol. Chemistry} 2009, {\bf 284}: 5457--5461.

\bibitem{Braunstein:2008bf}
Braunstein A, Mulet R, Pagnani A: {\bf Estimating the size of the solution space of metabolic networks.} {\it BMC Bioinformatics} 2008, {\bf 9:} 240.

\bibitem{vn}
Gale D. {\it The theory of linear economic models}. Chicago: The University of Chicago Press 1989.

\bibitem{dmmp}
De Martino A, Martelli C, Monasson R, Perez Castillo I:
{\bf Von Neumann's expanding model on random graphs.}
{\it J Stat Mech} 2007,  P05012.

\bibitem{Shulman2001a}
Shulman RG, Hyder F, Rothman DL:
{\bf Lactate efflux and the neuroenergetic basis of brain function.}
\newblock {\it NMR Biomed} 2001, {\bf 14}: 389-396.
\input{Shulman2001a}

\bibitem{Giove2003}
Giove F, Mangia S, Bianciardi M, Garreffa G, {Di Salle} F, Morrone R, Maraviglia B:
{\bf The physiology and metabolism of neuronal activation: in vivo studies by NMR and other methods.}
\newblock {\it Magn Reson Imaging} 2003, {\bf 21}: 1283-1293.
\input{Giove2003}

\bibitem{Shulman2011}
Shulman RG:
{\bf A Philosophical Analysis of Neuroenergetics.}
\newblock {\it Front Neuroenergetics} 2011,  {\bf 3}: 6.
\input{Shulman2011}

\bibitem{Raichle2002}
Raichle ME, Gusnard DA:
{\bf Appraising the brain's energy budget.}
\newblock {\it Proc Natl Acad Sci USA} 2002, {\bf 99}: 10237-10239
\input{Raichle2002}


\bibitem{Fox1986}
{Fox PT, Raichle ME}:
 {\bf {Focal physiological uncoupling of cerebral blood
  flow and oxidative metabolism during somatosensory stimulation in human
  subjects}.}
\newblock {{\it Proc Natl Acad Sci USA}} 1986, {\bf 83}: {1140-1144}.
\input{Fox1986}

\bibitem{Fox1988}
{Fox PT, Raichle ME, Mintun MA, Dence C}:
{\bf {Nonoxydative glucose
  consumption during focal physiologic neural activity}.}
\newblock {\it Science} 1988,  {\bf 241}: {462-464}.
\input{Fox1988}


\bibitem{Vafaee2012}
Vafaee MS, Vang K, Bergersen LH,  Gjedde A:
  {\bf Oxygen consumption and blood flow coupling in human motor cortex during intense finger tapping: implication for a role of lactate.}
\newblock {\it J Cereb Blood Flow Metab} 2012, {\bf 32}: 1859-1868.
\input{Vafaee2012}

\bibitem{Sibson1998}
Sibson NR, Dhankhar A, Mason GF, Rothman DL, Behar KL, Shulman RG:
  {\bf {S}toichiometric coupling of brain glucose metabolism and glutamatergic
  neuronal activity.}
\newblock {\it Proc Natl Acad Sci USA} 1998, {\bf 95}: 316--321.
\input{Sibson1998}


\bibitem{Sibson1997}
Sibson NR, Dhankhar A, Mason GF, Rothman DL, Behar KL, Shulman RG:
  {\bf {I}n vivo 13{C NMR} measurements of cerebral glutamine synthesis as evidence for glutamate-glutamine cycling.}
\newblock {\it Proc Natl Acad Sci USA} 1997, {\bf 94}: 2699-2704.
\input{Sibson1997}


\bibitem{Hyder2006}
Hyder F, Patel AB, Gjedde A, Rothman DL, Behar KL, Shulman RG:
  {\bf {N}euronal-glial glucose oxidation and glutamatergic-{GABA}ergic function.}
\newblock {\it J Cereb Blood Flow Metab} 2006, {\bf 26}: 865--877.
\input{Hyder2006}


\bibitem{Hertz2007}
Hertz L, Peng L, Dienel GA:
{\bf Energy metabolism in astrocytes: high rate of
  oxidative metabolism and spatiotemporal dependence on
  glycolysis/glycogenolysis.}
\newblock  {\it J Cereb Blood Flow Metab}  2007,  {\bf 27}: 219--249.
\input{Hertz2007}

\bibitem{yetanother}
Di Nuzzo M, Giove F:
{\bf Activity-dependent energy budget for neocortical signaling: effect of short-term synaptic plasticity on the energy expended by spiking and synaptic activity.} 
{\it J Neurosci Res} 2012, {\bf 90}: 2094--2102.

\bibitem{Shulman2001}
Shulman RG, Hyder F, Rothman DL:
{\bf Cerebral energetics and the glycogen
  shunt: neurochemical basis of functional imaging.}
\newblock {\it Proc Natl Acad Sci USA} 2001, {\bf 98}: 6417--6422.
\input{Shulman2001}

\bibitem{DiNuzzo2010d}
Di Nuzzo M, Mangia S, Maraviglia B, Giove F:
{\bf Glycogenolysis in astrocytes
  supports blood-borne glucose channeling not glycogen-derived lactate
  shuttling to neurons: evidence from mathematical modeling.}
\newblock {\it J Cereb Blood Flow Metab} 2010, {\bf 30}: 1895--1904.
\input{DiNuzzo2010d}

\bibitem{DiNuzzo2011}
{Di Nuzzo} M, Maraviglia B, Giove F:
{\bf Why does the brain (not) have
  glycogen?}
\newblock {\it Bioessays} 2011,  {\bf 33}: 319--326.
\input{DiNuzzo2011}

\bibitem{Pardo2011}
Pardo B, Rodrigues TB, Contreras L, Garzon M, Llorente-Folch I, Kobayashi K, Saheki T, Cerdan S, Satr\'{u}stegui J:
 {\bf  Brain glutamine synthesis requires neuronal-born aspartate as amino donor for
  glial glutamate formation.}
\newblock {\it J Cereb Blood Flow Metab} 2011, {\bf 31}: 90--101.
\input{Pardo2011}

\bibitem{Castro2009}
Castro MA, Beltr\'{a}n FA, Brauchi S, Concha II:
{\bf A metabolic switch in
  brain: glucose and lactate metabolism modulation by ascorbic acid.}
\newblock {\it J Neurochem} 2009,  {\bf 110}: 423--440.
\input{Castro2009}

\bibitem{Bak2009}
Bak L, Walls A, Schousboe A, Ring A, Sonnewald U, Waagepetersen HS:
{\bf {Neuronal
  glucose but not lactate utilization is positively correlated with
  NMDA-induced neurotransmission and fluctuations in cytosolic Ca2+ levels}.}
\newblock {\it J Neurochem} 2009, {\bf 109}: 87--93.
\input{Bak2009}

\bibitem{Bergersen2012}
Bergersen LH, Gjedde A:
{\bf Is lactate a volume transmitter of metabolic states of the brain?}
\newblock {\it Front Neuroenerg} 2012,  {\bf 4}:5. 
\input{Bergersen2012}

\bibitem{Gordon2008}
Gordon GRJ, Choi HB, Rungta RL, Ellis-Davies GCR, MacVicar BA:
{\bf Brain
  metabolism dictates the polarity of astrocyte control over arterioles.}
\newblock {\it Nature} 2008,  {\bf 456}: 745--749.
\input{Gordon2008}

\bibitem{Shimizu2007}
Shimizu H, Watanabe E, Hiyama TY, Nagakura A, Fujikawa A, Okado H, Yanagawa Y, Obata K, Noda M:
{\bf Glial
  Nax channels control lactate signaling to neurons for brain [Na$^+$] sensing.}
\newblock {\it Neuron} 2007,  {\bf 54}: 59--72.
\input{Shimizu2007}

\end{thebibliography}

\begin{thebibliography}{}

\bibitem{Jamshidi:2001kl}
Jamshidi, Edwards J, Fahland T, Church B, Palsson B (2001) Dynamic simulation of the human red blood cell metabolic network. Bioinformatics 17: 286--287

\bibitem{Varma:1994kc}
Varma A, PalssonB (1994) Metabolic Flux Balancing: Basic Concepts, Scientific and Practical Use. Nature Biotech 12: 994--998

\bibitem{kau}
Kauffman K, Prakash P, Edwards J (2003) Advances in flux balance analysis. Curr Opin Biotech 14: 491--496

\bibitem{Lee:2006qo}
Lee JM, Gianchandani E, Papin J (2006) Flux balance analysis in the era of metabolomics. Brief Bioinf 7: 140--150

\bibitem{palsson}
Palsson B (2006) {\it Systems biology: Properties of reconstructed networks} (Cambridge University Press)

\bibitem{Beard:2008fk}
Beard DA, Qian H (2008) {\it Chemical Biophysics: Quantitative Analysis of Cellular Systems} (Cambridge Univesity Press)

\bibitem{Oberhardt:2009gb}
Oberhardt M, Chavali A, Papin J (2009) Flux balance analysis: interrogating genome-scale metabolic networks. Meth Mol Biol 500: 61--80

\bibitem{Orth:2010if}
Orth J, Thiele I, Palsson B (2010) What is flux balance analysis? Nature Biotech 28: 245--248

\bibitem{Price:2004lp}
Price N, Schellenberger J and Palsson B (2004) Uniform Sampling of Steady-State Flux Spaces: Means to Design Experiments and to Interpret Enzymopathies. Biophys J 87: 2172--2186

\bibitem{Braunstein:2008bf}
Braunstein A, Mulet R, Pagnani A (2008) Estimating the size of the solution space of metabolic networks. BMC Bioinformatics 9: 240

\bibitem{Varma:1994gd}
Varma A, Palsson B (1994) Stoichiometric flux balance models quantitatively predict growth and metabolic by-product secretion in wild-type Escherichia coli W3110. Appl Environ Microbiol 60: 3724--3731

\bibitem{Edwards:2002bd}
Edwards J, Covert M, Palsson B (2002) Metabolic modelling of microbes: the flux-balance approach. Environ Microbiol 4: 133--140

\bibitem{Feist:2008bs}
Feist A, Palsson B (2008) The growing scope of applications of genome-scale metabolic reconstructions using \emph{Escherichia coli}. Nature Biotech 26: 659--667

\bibitem{Lewis:2010fu}
Lewis NE, Hixson KK, Conrad TM, Lerman JA, Charusanti P, Polpitiya AD, Adkins JN, Schramm G, Purvine SO, Lopez-Ferrer D, Weitz KK, Eils R, K\"onig R, Smith RD, Palsson B (2010) Omic data from evolved E. coli are consistent with computed optimal growth from genome-scale models. Mol Sys Biol 6: 390

\bibitem{gale}
Gale D (1960) {\it The theory of linear economic models} (The University of Chicago Press)

\bibitem{Imielinski:2005dz}
Imielinski M, Belta C, Halasz A, Rubin H (2005) Investigating metabolite essentiality through genome-scale analysis of Escherichia coli production capabilities. Bioinformatics 21: 2008--2016

\bibitem{ib00}
Imielinski M, Belta C, Rubin H, Halasz A (2006) Systematic Analysis of Conservation Relations in Escherichia coli Genome-Scale Metabolic Network Reveals Novel Growth Media. Biophys J 90: 2659--2672

\bibitem{Warren:2007fk}
Warren PB, Jones JL (2007) Duality, thermodynamics, and the linear programming problem in constraint-based models of metabolism. Phys Rev Lett 99: 108101

\bibitem{ploscb}
De Martino D, Figliuzzi M, De Martino A, Marinari E (2012) A Scalable Algorithm to Explore the Gibbs Energy Landscape of Genome-Scale Metabolic Networks. PLoS Comput Biol 8(6): e1002562.

\bibitem{Martino:2005mi}
{De Martino} A, Marsili M (2005) Typical properties of optimal growth in the Von Neumann expanding model for large random economies. J Stat Mech L09003

\bibitem{Martino:2007zt}
{De Martino} A, Martelli C, Monasson R, Perez Castillo I (2007) Von Neumann's expanding model on random graphs. J Stat Mech P05012

\bibitem{Martino:2009lh}
{De Martino} A, Martelli C, Massucci FA (2009) On the role of conserved moieties in shaping the robustness and production capabilities of reaction networks. Europhys Lett 85: 38007

\bibitem{Martino:2010tw}
{De Martino} A, Figliuzzi M and Marsili M (2010) One way to grow, many ways to shrink: the reversible Von Neumann expanding model. J Stat Mech P07032

\bibitem{linopt}
Schrijver A (1986) {\it Theory of linear and integer programming}. (John Wiley \& Sons Ltd)

\bibitem{minover0}
Krauth W, Mezard M (1987) Learning algorithms with optimal stability in neural networks. J Phys A 20: L745--L752

\bibitem{Martelli:2009jl}
Martelli C, {De Martino} A, Marinari E, Marsili M, Perez Castillo I (2009) Identifying essential genes in Escherichia coli from a metabolic optimization principle. Proc Nat Acad Sci USA 106: 2607--2611

\bibitem{granata}
{De Martino} A, Granata D, Marinari E, Martelli C, Van Kerrebroeck V (2010) Optimal Fluxes, Reaction Replaceability, and Response to Enzymopathies in the Human Red Blood Cell. J Biomed Biotech 2010: 415148

\bibitem{Kyoto10}
{De Martino} A, Marinari E (2010) The solution space of metabolic networks: producibility, robustness and fluctuations. J Phys (Conf Ser) 233: 012019

\end{thebibliography}
\end{document}